\documentclass[manuscript]{aastex}
\usepackage{natbib}
\usepackage{color}

\slugcomment{Accepted for publication in ApJ}

\shorttitle{Tracing the Outflow of a z=0.334 FeLoBAL}
\shortauthors{Lucy et al.}

\begin{document}


\title{Tracing the Outflow of a z=0.334 FeLoBAL: New Constraints from
  Low-Ionization Absorbers in FBQS J1151+3822}


\author{Adrian B. Lucy and Karen M. Leighly\footnote{\footnotesize Visiting
    Astronomer, Kitt Peak National Observatory, which is operated by
    the Association of Universities for Research in Astronomy (AURA)
    under cooperative agreement with the National Science Foundation.}}
\affil{Homer L.\ Dodge Department of Physics and Astronomy, The
  University of Oklahoma, 440 W.\ Brooks St., Norman, OK 73019}

\author{Donald M. Terndrup$^1$}
\affil{Department of Astronomy, The Ohio State University, 4055
  McPherson Laboratory, 140 W.\ 18th Ave., Columbus, OH 43210}

\author{Matthias Dietrich$^1$}
\affil{Department of Physics and Astronomy, Ohio University,
  Clippinger Labs 251B, Athens, OH 45701}

\and

\author{Sarah C. Gallagher}
\affil{Department of Physics \& Astronomy, The University of Western
  Ontario, London, ON, N6A 3K7, Canada}



\begin{abstract}

We show for the first time that FBQS~J1151$+$3822 is an iron low-ionization broad absorption line quasar (FeLoBAL QSO), the second-brightest and second-closest known example of this class. \ion{He}{1}* and \ion{Fe}{2} together act as an effective analytical tool, allowing us to obtain useful kinematic constraints from photoionization models of the outflow without needing to assume any particular acceleration model. The main outflow's log ionization parameter is $-1.5$, the log hydrogen density ($\rm cm^{-3}$) 5.5--8, the log hydrogen column density ($\rm cm^{-2}$) 21.7--21.9, the absorption radius 7.2--$127\rm\, pc$, and the kinetic luminosity 0.16\%--4.5\% of the bolometric luminosity. We obtain line-of-sight covering fractions of $\sim 0.25$ for strong \ion{Fe}{2}, $\sim 0.5$ for \ion{He}{1}*, and $\sim 0.6$ for \ion{Mg}{2}. Narrower and shallower absorption lines from weaker \ion{Fe}{2} and \ion{Mn}{2} with outflow velocity $\sim 3400\rm \,km\, s^{-1}$ have appeared between 2005 and 2011, suggesting that dense cores may have condensed inside the main outflow. Consideration of the literature might suggest that the FBQS~J1151$+$3822 outflow is a member of a rare and distinct subclass of FeLoBALs with high densities and correspondingly small absorption radii. We find, however, that such outflows are not necessarily a distinct subclass, and that their apparent rarity could be a symptom of selection bias in studies using density-sensitive lines.

\end{abstract}





\section{Introduction\label{intro}}

A substantial fraction of quasars exhibit broad, blue-shifted
absorption lines in their rest-frame UV spectra: manifestations of
tempestuous winds, emerging from the central engine at up to thousands
of kilometers per second \citep{weymann91}.  These
broad absorption line quasar (BALQSO) outflows can distribute
chemically enriched gas through the intergalactic medium
\citep{cavaliere02}, facilitate accretion by carrying away angular
momentum \citep{emmering92, konigl94}, and inject kinetic energy into the host galaxy, perhaps suppressing star formation
\citep{so04, farrah12}. As such, determining their structure and
kinetic power is essential for understanding the physics of
supermassive black hole accretion disks and for discriminating between
different models of galaxy evolution.

BALQSOs have been divided into three classes based on their spectral
properties. High-ionization BALQSOs (HiBALs) have absorption
lines from high-ionization species such as \ion{C}{4}. Low-ionization
BALQSOs (LoBALs) have the same high-ionization lines as HiBALs, but
also have absorption from lower-ionization species such as \ion{Mg}{2}
\citep{voit93}. Finally, FeLoBALs have the same lines as
LoBALs, but also have absorption from \ion{Fe}{2} \citep{hazard87,
  becker97}.  HiBALs are the most common type of BALQSO, comprising perhaps 10--26\% of quasars, while FeLoBALs are the least
common type, comprising at most $\sim 1$\% of quasars
\citep{trump06,gibson09}. As we will argue in the course of our photoionization analyses, the appropriate physical interpretation of BALQSO subclasses relates to the column density of the outflowing gas relative to the
size of the \ion{H}{2} region; for a given ionization parameter, HiBAL outflows have the smallest column, LoBAL outflows have a larger column (but \ion{Mg}{2} LoBALs can still be truncated before the hydrogen ionization front is breached), and FeLoBAL outflows have the largest column (thicker, at least marginally, than the \ion{H}{2} zone). 

FeLoBALs are of particular interest, in part as a flashpoint in the broader debate over BALQSO unification. There are two limiting models used to explain the rarity of LoBALs relative to HiBALs, and the more dramatic rarity of FeLoBALs: an orientation model wherein the lines of sight through which a high column would be observed subtend a small solid angle \citep[see, e.g.,][for supporting evidence]{green01, gallagher06, morabito11}, and an evolutionary model wherein LoBALs and FeLoBALs are a short-lived stage in the lifetime of a young quasar outflow dispersing its shroud of dust and gas \citep{voit93, becker00, gregg06} and suppressing  \citep{farrah12} high star formation rates \citep{farrah07, cs01}. This dichotomy of theory motivates detailed investigation into the geometry and kinetic power of FeLoBALs, the rarest class of BALQSO.

In 2008, we observed the nearby ($z=0.334$), luminous ($M_V=-25.5$)
quasar FBQS~J1151$+$3822 using the NASA Infrared Telescope Facility
(IRTF), and presented our results in \citet{leighly11}.  We serendipitously
discovered a 
\ion{He}{1}*$\lambda 10830$ (2s$\rightarrow$2p) BAL.  The \ion{He}{1}*$\lambda 3889$ (2s$\rightarrow$3p)
BAL has been seen in a number of quasars and utilized to
great effect \citep{korista08,arav08,dunn10,borguet12}, but
this was the first time that the 10830\AA\/ component had been
reported. In that paper, in order to obtain useful (i.e., to within an order of magnitude or so) parameter constraints (on, e.g., ionization parameter, density, column density, radius, and kinetic luminosity) from the limited information afforded by \ion{He}{1}* alone, we assumed acceleration by radiative line driving---one of several possible accelerating mechanisms, as discussed in \S4.2 of that paper, whereby
photons from the central engine resonantly scatter with atoms in the
outflow and thereby impart their momentum to the flow. Through a dynamical argument presented in \S4.2 of that paper, this assumption yielded a lower limit on the density, without which our absorption radius and kinetic luminosity constraints spanned many orders of magnitude.

In 2011 May, we observed FBQS~J1151$+$3822 using the KPNO 4-meter
telescope equipped with the RC spectrograph, covering 2371-4340\AA\/ in the rest frame.  Our spectrum, reduced in
\S\ref{obs}, reveals for the first time that FBQS~J1151$+$3822 is an FeLoBAL, featuring \ion{Fe}{2} and \ion{Mg}{2} BALs in the previously unobserved wavelengths short of $\sim 2845$\AA\/. Interested in how these lines could complement the information gleaned from our prior analysis of \ion{He}{1}* lines in this object, we performed
phenomenological spectral fits in \S\ref{analysis}, with template analysis based on the
\ion{He}{1}* lines. {\it Cloudy}
photoionization simulations and partial covering analysis, leading to
physical spectral models, are described in \S\ref{cloudy}. In that section, we extrapolate constraints on the geometry and kinetic parameters of the outflow, with \ion{Fe}{2} allowing us to drop our assumption of acceleration by radiative line driving; \ion{He}{1}* and \ion{Fe}{2} together amount to a powerful analytical tool, with \ion{He}{1}* pinpointing the ionization parameter, and \ion{Fe}{2} (along with other once-ionized iron-peak elements) bracketing the hydrogen column range.

We note that the present paper
incorporates a significant amendment to our prior analysis of
\ion{He}{1}* lines; we now include a model of the host galaxy's
contribution to the $\sim 1$ micron continuum, described in Appendix
\ref{host_gal}, which alters the \ion{He}{1}* covering fraction but does not affect any other results from \citet{leighly11}.

In \S\ref{inhomogeneous}, we note that this type of analysis, with enough ions, could potentially obtain the distribution of hydrogen column as a function of covering fraction and build a model of inhomogeneity in the outflow. In \S\ref{comparison}, we compare our target to other FeLoBALs. In \S\ref{den_sen}, we argue that the relatively small radius we obtain for the FBQS~J1151$+$3822 outflow is not anomalous or extreme, but only appears to be so due to selection biases in much of the existing literature; we caution against premature assertions regarding the supposed near-ubiquity of large absorption radii among BAL outflows. A summary is presented in \S\ref{summary}.

We used cosmological parameters $\Omega_\Lambda$=0.73,
$\Omega_M=0.27$, and $H_0=71\rm\, km\, s^{-1}\, Mpc^{-1}$, unless
otherwise specified.

\section{Observation and Data Reduction\label{obs}}

Our spectra were taken during a 2011 May 6--10 observing run at Kitt Peak National
Observatory on the Mayall 4-meter telescope.  We
used the KPC--007 grism (dispersion of 1.39\AA\/ per pixel, resolution of 3.5\AA\/, and one of the bluer gratings
available, with good sensitivity up to the atmospheric cutoff), the t2ka CCD (gain of 1.4).  We observed
FBQS~J1151$+$3822 using a blue setting with nominal wavelength
coverage between $\sim 3200$--6000\AA\/.  The sensitivity
dropped steeply at both ends due to vignetting and, especially, due to
the atmospheric cutoff.    

Six 15-minute observations were performed, for a total exposure of 90
minutes, as the target passed through the meridian.  Atmospheric conditions were not photometric, and
the brightness of individual exposures varied by a factor of $\sim
1.33$.   Interestingly, the KPNO spectrum, though it was taken during non-photometric conditions, and the Sloan Digital Sky Survey (SDSS) spectrum (observed 2005 May 11) are almost exactly equal
in flux in those regions where they overlap, perhaps suggesting that the target
has brightened.  

\begin{figure}[h]
\epsscale{1.0}
\includegraphics[width=6.5truein]{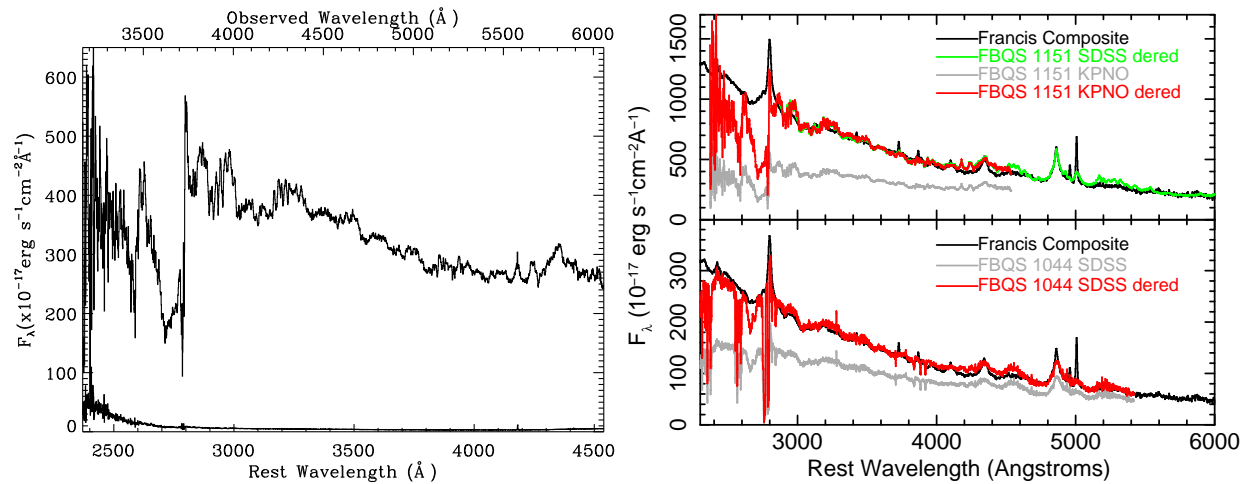}
\caption{\footnotesize The KPNO Mayall 4-meter telescope spectrum of FBQS~J1151$+$3822, as described in \S\ref{obs}.
  The left panel shows the observed spectrum and the assigned uncertainties.  The right panel shows the 
  dereddening for FBQS~J1151$+$3822 and
  for a  comparison BALQSO FBQS~J1044$+$3656. We find that the FBQS~J1151$+$3822 continuum is consistent with the Francis composite when dereddened by $E(B-V)=0.14$ using an SMC reddening curve. \label{fig1}}
\end{figure}

We performed a standard reduction using IRAF. 
A wide range of flux standard stars were observed; however, the
spectra from some of these stars showed peculiar decreases in flux at
the ends of the spectra that we believe were caused by the stars not
being centered well in the slit.   Therefore, only stars observed at
the lowest air masses were used for flux calibration. The spectra
were combined, weighted by the flux between 4000 -- 5500\AA\/. The Galactic reddening in the direction of FBQS~J1151$+$3822 is 
$E(B-V)=0.023$ \citep{sfd98}, which we corrected for using the CCM  \citep{ccm89}
reddening curve. The
NED\footnote{\footnotesize NASA/IPAC   Extragalactic Database;
  http://ned.ipac.caltech.edu/} redshift was listed as 0.3344, and we
used that value to correct for cosmological expansion. IRAF spectra do not include errors; FBQS~J1151$+$3822 is a bright object, so we gauged 1-$\sigma$ errors by assuming Poisson noise in the uncalibrated photon count spectrum. The reduced spectrum and
assigned errors are shown in Fig.~\ref{fig1}.

In \citet{leighly11}, we inferred that FBQS~J1151$+$3822 was intrinsically reddened,
based on a comparison of the SDSS/MDM spectra and SDSS photometry,
with the mean quasar spectral energy distribution (SED) published by
\citet{richards06}. We dereddened by $E(B-V)=0.1$ using
a Small Magellanic Cloud (SMC) reddening law \citep{pei92}.  For the present work, we addressed the question of
reddening more carefully by comparing the FBQS~J1151$+$3822 spectrum
with several quasar composite spectra.   We found the best match with
the updated \citet{francis91} quasar composite spectrum created using
Large Bright Quasar Sample (LBQS) spectra\footnote{\footnotesize Available at www.mso.anu.edu.au/pfrancis/composite/widecomp.d.}.  We again
used an SMC reddening curve, and determined that FBQS~J1151$+$3822
best matches the quasar composite if $E(B-V)=0.14$. Interestingly, this is the typical value of reddening found for LoBALs by \citet{gibson09}. In Fig.~\ref{fig1}, we show the scaled
composite spectrum overlaid with the original and dereddened
FBQS~J1151$+$3822 spectra.  For comparison, we also show the same for
the SDSS spectrum of the $z=0.7$ broad absorption line quasar
FBQS~J1044$+$3656; for that object, we inferred a reddening of
$E(B-V)=0.09$.  After dereddening of FBQS~J1151$+$3822, the inferred apparent magnitude is
$m_V=15.4$, and the inferred absolute magnitude is $M_V=-25.8$. 

In fact, FBQS~J1151$+$3822 most likely would have been included among
the PG quasars were it not for its BAL-related properties (absorption lines in the U band, together with intrinsic reddening).  The SDSS photometry predicts
$U-B=-0.42$, above the nominal PG cutoff of $U-B=-0.44$, and $B=16.0$, just consistent with the corresponding nominal PG cutoff \citep{jester05}. Correcting for emission and absorption lines (as in Appendix~\ref{host_gal}) predicts $U-B=-0.64$ and $B=15.96$.  Correcting for the
inferred $E(B-V)=0.14$ reddening intrinsic to the quasar as
well predicts $U-B=-0.69$ and $B=15.81$.  \citet{jester05} notes that
the effective (median) color cutoff for the PG sample is $U-B=-0.71$. Intrinsically,
FBQS~J1151$+$3822 is as bright as a PG quasar.

\section{Phenomenological Analysis\label{analysis}}

In \citet{leighly11}, we described the metastable \ion{He}{1}* 
absorption in FBQS~J1151$+$3822 that was observed both in the infrared
at 10830\AA\/ and in the optical at 3889\AA\/.   However, the optical spectra
presented in that paper covered a rest-frame wavelength range of only
2845--6910\AA\/ (SDSS) and 3204--4401\AA\/ (MDM).  The new KPNO
spectrum spans 2371--4340\AA\/ in the rest frame, revealing large absorption complexes
shortward of \ion{Mg}{2}$\lambda 2800$, analyzed in \S\ref{prelim}
and \S\ref{abs_modeling}. In addition, new absorption lines have
appeared just short of 2960\AA\/, which we describe in
\S\ref{new_lines}. 

\subsection{Motivation for Template Analysis\label{prelim}}

\begin{figure}[h]
\epsscale{1.0}
\includegraphics[width=6.5in]{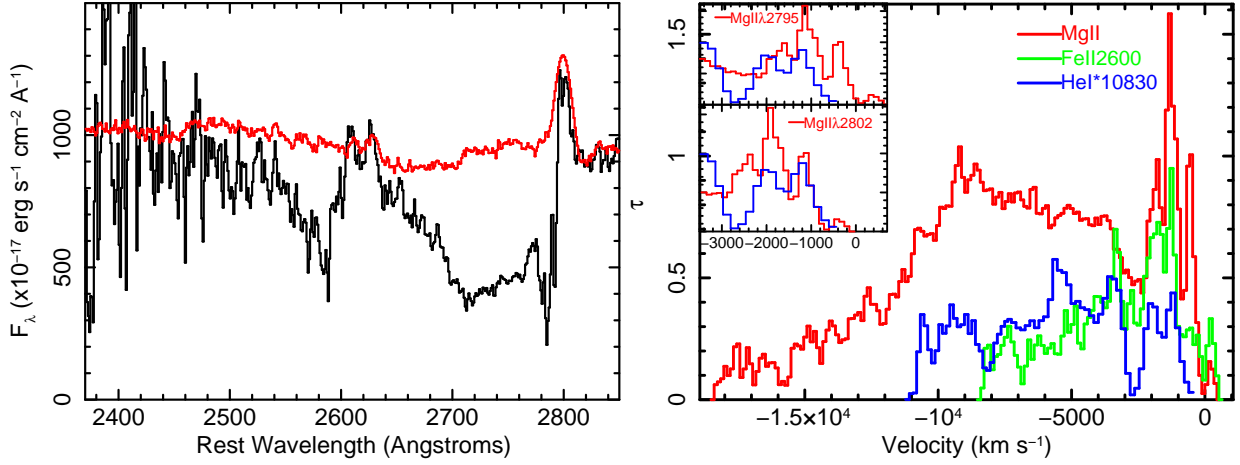}
\caption{\footnotesize Preliminary continuum and apparent optical depths for the
  absorption lines.  The left panel shows the dereddened
  FBQS~J1151$+$3822 spectrum overlaid with the continuum model
  developed from our PG~1543$+$489 spectrum.  The right panel shows the
inferred apparent optical depths for the feature spanning $\sim 2650$
to $\sim 2800$\AA\/ (in red), if the whole feature is attributed to
the \ion{Mg}{2} doublet, and the feature between $\sim 2540$ and $\sim
2600$\AA\/ (green) if 
attributed to the strong \ion{Fe}{2} ground-state transition at
2600\AA\/, compared with the  \ion{He}{1}*$\lambda 10830$ apparent
optical depth profile (blue) derived in \citet{leighly11}. Significant
structural similarities between these profiles are observed, while the
breadth of the \ion{Mg}{2} suggests contamination by excited-state
\ion{Fe}{2} on its blue wing.  The insets show the low-velocity features from the rest wavelengths of the individual lines forming the \ion{Mg}{2}
doublet.\label{fig2} } 
\end{figure}

To investigate the origin of absorption, we first needed to determine
the level of the unabsorbed continuum ($I_0$).  In \citet{leighly11},
we noted that the \ion{Fe}{2} emission in PG~1543$+$489 near \ion{He}{1}*$\lambda 3889$ was similar to that of
FBQS~J1151$+$3822, and we
initially thought to use a spectrum of PG~1543$+$489 observed at KPNO the same night.  The continuum in Fig.~\ref{fig2} is a model of said spectrum with \ion{Mg}{2} emission scaled to match that of
FBQS~J1151$+$3822.  

There are two clearly distinct regions of absorption short of
2800\AA\/. At least some of the absorption between $\sim 2650$ and
$2800$\AA\/ must be due to \ion{Mg}{2}, so the right panel of
Fig.~\ref{fig2} plots apparent optical depth for this region as a
function of velocity from the oscillator-strength-weighted mean rest
wavelength of the \ion{Mg}{2}$\lambda 2800$ doublet, using
$I/I_0(v)=exp(-\tau(v))$. The strongest of the many lines possible in
the other absorption region (predominately transitions from the ground
term of \ion{Fe}{2}) would absorb from ground state \ion{Fe}{2} at
$2600$\AA\/, so we likewise plot the apparent optical depth between
$\sim 2540$ and $2600$\AA\/ as a function of velocity from
2600\AA\/. For comparison, we overlay the apparent optical depth of
the \ion{He}{1}*$\lambda 10830$ line reported in \citet{leighly11}.

This plot reveals several salient properties of the absorption shape
between $\sim 2650$\AA\/ and 2800\AA\/.  Most obviously, it is significantly
broader than the \ion{He}{1}* absorption trough.  \ion{He}{1}* is
absorption from a neutral atom, but because metastable helium arises
from recombination onto He$^+$, this absorption occurs in roughly the
same gas that would produce the generally-prominent \ion{C}{4}
BAL. That is, \ion{He}{1}* acts as a high-ionization line, while
\ion{Mg}{2} is a low-ionization line. High-ionization lines are
generally broader, and extend to higher velocities, than
low-ionization lines \citep{voit93}; if this feature
originated in \ion{Mg}{2} only, its breadth relative to \ion{He}{1}*
would be very anomalous. Excited-state \ion{Fe}{2}, then, is the most
plausible origin for the short-wavelength end of this feature; we show
in \S\ref{upper_limit} that \ion{Cr}{2} and \ion{Ti}{2}, the only
other conceivable absorbers on the blue wing of \ion{Mg}{2}, can at
most produce a small fraction of the observed opacity.

There are several narrow absorption features near 2800 \AA\/, including those at $\sim -1870$,
$-1260$, $-580$, and $+210\rm \, km\, s^{-1}$ from the mean rest
wavelength of the \ion{Mg}{2} doublet.  The velocity separation of the
\ion{Mg}{2} doublet is $769\rm \, km\, s^{-1}$, and the two
lowest-velocity features, which are relatively narrow, have this
separation. They thus appear to be low-velocity associated absorption
with $v\sim 390\rm \, km\, s^{-1}$, unrelated to the BAL outflow, and
we interpolate over them for fits throughout this paper. The features at $\sim -1870$ and $\sim -1260\rm \, km\,s^{-1}$ are
separated by only $\sim 610\rm \, km\, s^{-1}$, and thus cannot be
attributed to the doublet structure of \ion{Mg}{2}.  These features
are echoed in the \ion{Fe}{2} opacity shortward of 2600\AA\/. The
\ion{He}{1}* profile, similarly, has two low-velocity components;
however, as shown in Fig.~\ref{fig2}, these do not line up perfectly
with the KPNO spectrum features.

Despite the differences, we suggest that the profile of \ion{He}{1}*
is more similar to those of \ion{Mg}{2} and \ion{Fe}{2} than it is
different; in particular, there is a suggestive low-opacity region in
all said profiles at $\sim2700 \rm km\, s^{-1}$.  Due to significant
line blending within \ion{Fe}{2} multiplets in the bandpass of the
KPNO spectrum, and the blending of \ion{Mg}{2} with excited-state
\ion{Fe}{2} as noted above, we could not hope to deconvolve the line
contributions and extract ionic columns without using some non-blended
absorption line as a template.  Blending also precluded
modeling individual velocity components of the outflow as independent
gases. We therefore used the \ion{He}{1}* profiles as templates in the following analyses.

\subsection{The Continuum\label{other_cont}}

While the \ion{Fe}{2} emission in PG~1543$+$489 resembles that of
FBQS~J1151$+$3822, our KPNO spectrum shows dramatic differences
in multiplet strength at shorter wavelengths. UV 
\ion{Fe}{2} emission complexes are known to display a variety of shapes
\citep[e.g.,][]{lm06, leighly07}. Therefore, we created customized continua for
FBQS~J1151$+$3822.  This object has especially strong and blocky
\ion{Fe}{2} features between 2900 and 3000\AA\/.  We used spectral fit results from an unpublished
catalog of 5307 SDSS spectra (chosen because it had enough
spectra with emission structures matching our object)  in the
z=1.2--1.8 range, selecting quasars with relatively narrow \ion{Mg}{2}
emission lines (to match that of FBQS~J1151$+$3822) and no broad
absorption lines (to reveal the continuum).  We identified a set of
objects that had similarly large, blocky $\sim 2950$\AA\/ features. 

\begin{figure}[p]
\epsscale{1.0}
\includegraphics[width=6.5in]{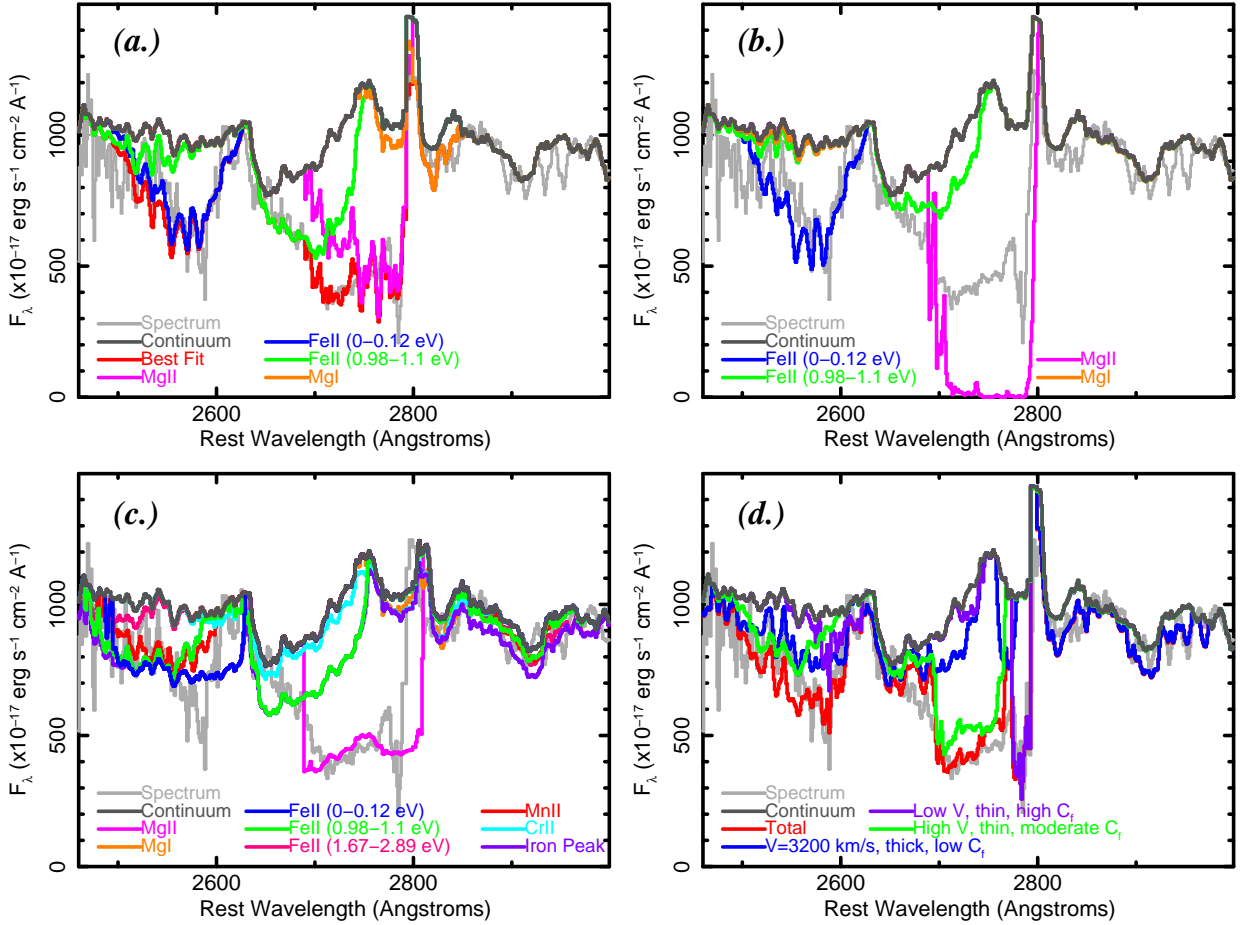}
\caption{\footnotesize  
    (a.) The phenomenological model fit, discussed in \S\ref{analysis}. High-excitation \ion{Fe}{2} is necessary to fit the spectrum just short of \ion{Mg}{2}, a region which anchors the high-excitation \ion{Fe}{2} fit. (b.)  The model spectrum from the
  figure-of-merit best fit obtained through comparison of the columns
  obtained through the phenomenological fit and {\it Cloudy} modeling, as
  described in \S\ref{fom}. This model does not match observations because it does not take partial covering and saturation into account. (c.)  The model spectrum based on the upper limit analysis
  described in  \S\ref{upper_limit}, for a covering fraction of 0.58
  applied to \ion{Mg}{2} and \ion{Mg}{1}, and a covering fraction of 0.25 applied to
  \ion{Fe}{2} and other iron-peak ions (\ion{Mn}{2}, \ion{Cr}{2}). (d.) An ad hoc inhomogeneous model, described in
  \S\ref{plausible}, that attempts to explain the $\sim 2960$\AA\/ lines without producing excess absorption elsewhere. \label{FIG3}} 
\end{figure}

After experimenting with several composite continua, we settled on one constructed
from 12 objects with large iron equivalent widths (between 50 and 180\AA\/)
and with narrow \ion{Mg}{2} ($<2400\rm\, km\, s^{-1}$).  This is the continuum model appearing in every panel of Fig.~\ref{FIG3}. To estimate the systematic uncertainties associated with our choice of continuum, we fit our spectrum with this continuum, the PG~1543+489
continuum, and one other constructed using quasars from the above catalog with the strongest \ion{Fe}{2}. As reported in \S\ref{fit_results}, continuum-related systematic uncertainty in fit results is relatively small.

\subsection{Modeling the Apparent Absorption Short of 2852 \AA\/ \label{abs_modeling}}

Here, we describe our extraction of the apparent column densities of ions absorbing between $2481$ (where the statistical error in our spectrum stabilizes into a smooth descent) and $2852$\AA\/ (the rest wavelength of a possible \ion{Mg}{1} BAL). To achieve this extraction, we made the assumptions---the utility of which will become evident via \S\ref{cloudy} photoionization models---that all opacity fully and homogeneously covers the continuum, and is produced by the strongest BALQSO absorbers in this region: \ion{Fe}{2} lines from less than 2.5~eV above the ground state, the \ion{Mg}{2}$\lambda 2800$ doublet, and perhaps the \ion{Mg}{1}$\lambda 2852$ line. 

\subsubsection{\ion{Fe}{2} Transitions Considered\label{transitions}}

\begin{figure}[h]
\epsscale{1.0}
\includegraphics[width=6.5in]{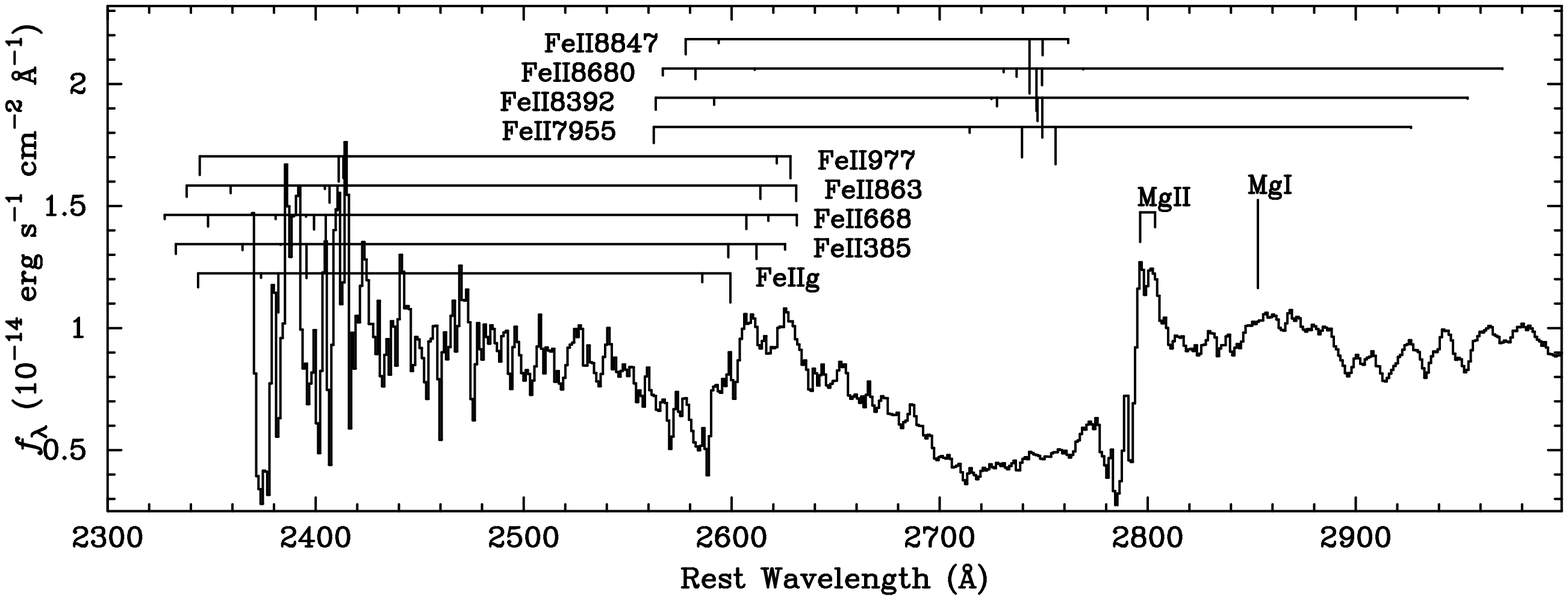}
\caption{\footnotesize The transitions considered in \S\ref{abs_modeling}, along with term-related shorter-wavelength lines that fall outside our band of good data.
  \ion{Fe}{2} lines are grouped by common lower level energy (labeled in $\rm cm^{-1}$) rather
  than by multiplet.  Only lines with $f_{ik}> 0.005$ are shown.
  The length of the downward tick mark is proportional to the
  oscillator strength of the transition with, for visibility's sake, the scaling of
  \ion{Fe}{2} being 2.5 times that of \ion{Mg}{2} and \ion{Mg}{1}. The iron lines group nicely by wavelength and excitation, with the bottom five energy levels (\ion{Fe}{2}\,Low) dominating short of $\sim 2600$\AA\/, and the top four energy levels (\ion{Fe}{2}\,High) dominating on the blue wing of \ion{Mg}{2}. The observed spectrum towards $\sim 2500$ \AA\/  and shortward is dominated by statistical noise due to the atmospheric cutoff, so any structure there is not real.\label{FIG4}}
\end{figure}

Between 2481 and 2852\AA\/, Fe$^+$ BAL absorption is dominated by four multiplets absorbing from within 2.5 eV of the ground state; these lines are summarized in Table~\ref{tbl-nom} and superimposed on the spectrum in Fig.~\ref{FIG4}, using NIST\footnote{\footnotesize National Institute of Standards and Technology Atomic Spectra Database; http://www.nist.gov/pml/data/asd.cfm} atomic data.  The upshot of this information, when considered in light of the two
distinct absorption regions previously discussed in \S\ref{prelim}, is that
\ion{Fe}{2} absorption can be split into two sets, with distinct and
narrow energy ranges, that are useful for template analysis
fits. Absorption from the a\,$^6$D term (0--0.12~eV above ground) dominates the absorption
complex short of $\sim 2600$\AA\/. Absorption from a\,$^4$D
(0.98--1.1~eV) is solely responsible for the blue wing of the
\ion{Mg}{2} complex, and also contributes to the complex short of
$\sim 2600$\AA\/. For the convenience of readers unfamiliar with the
atomic structure of Fe$^+$, we will henceforth call the a\,$^6$D set
of lines ``\ion{Fe}{2}\,Low," and the more excited a\,$^4$D set
``\ion{Fe}{2}\,High." 

\begin{deluxetable}{lccccr}
\tablewidth{0pt}
\tablecaption{\ion{Fe}{2} Line Sets\tablenotemark{~}}
\tablehead{
 \colhead{Name} &
 \colhead{Lower-level Excitation (eV)} &
 \colhead{Lower-level Term} &
 \colhead{Lines} &
 \colhead{Multiplets} &
 \colhead{Rest $\lambda$ (\AA\/)}\label{tbl-nom}
}
\startdata
\ion{Fe}{2}\,High &  0.98--1.1 & a\,$^4$D & 19 & 2 &  2692--2773 \\
& & & 8 & 1 & 2562--2612 \\
\ion{Fe}{2}\,Low & 0--0.12 & a\,$^6$D & 13 & 1 & 2585--2631 \\
\enddata
\label{table6}
\tablenotetext{~}{\footnotesize{This information is comprehensive only for this bandpass, and is not comprehensive for extremely saturated BALs where \ion{Fe}{2} absorbs from above 2.5 eV (as in \S\ref{upper_limit}). We also omit multiplets with line oscillator strengths less than $\sim 0.001$.}
}
\end{deluxetable}

\subsubsection{Template Analysis\label{profile}}

To facilitate a fit of the apparent optical depth that could produce
useful results in spite of line blending, we
first constructed amalgamated optical depth profiles for \ion{Mg}{2},
\ion{Fe}{2}\,Low, and \ion{Fe}{2}\,High. We used the unblended
\ion{He}{1}* lines as templates to shape these profiles. \citet{korista92} and \citet{arav99}, among others, have previously used templates to de-blend BALs.

Within, for example, the doublet \ion{Mg}{2} line set, we weighted
each component transition's contribution to the total optical depth of
the set as
\begin{equation}
\tau_{j,proto}(v)=R_i \frac{\lambda_j f_j}{\lambda_{He} f_{He}}
\tau_{He}(v).
\label{equation1}
\end{equation}
Here, $\tau_{j,proto}$ denotes a weighted optical depth for a
transition $j$ with lower energy level $i$, as a function of velocity
$v$. $\tau_{He}(v)$ is the \ion{He}{1}* profile used as a
template. For the simple case of \ion{Mg}{2}, where both doublet lines
arise from the same lower level $i$, $R_i=1$ and the relative strength
of each line $j$ depends only on the product of the wavelength and
oscillator strength of the line, $\lambda_j f_j$
\citep{ss91}. For computational convenience, we also
divided by the \ion{He}{1}* template line's $
\lambda_{He} f_{He}$.

\begin{figure}[h]
\epsscale{1.0}
\includegraphics[width=6.5in]{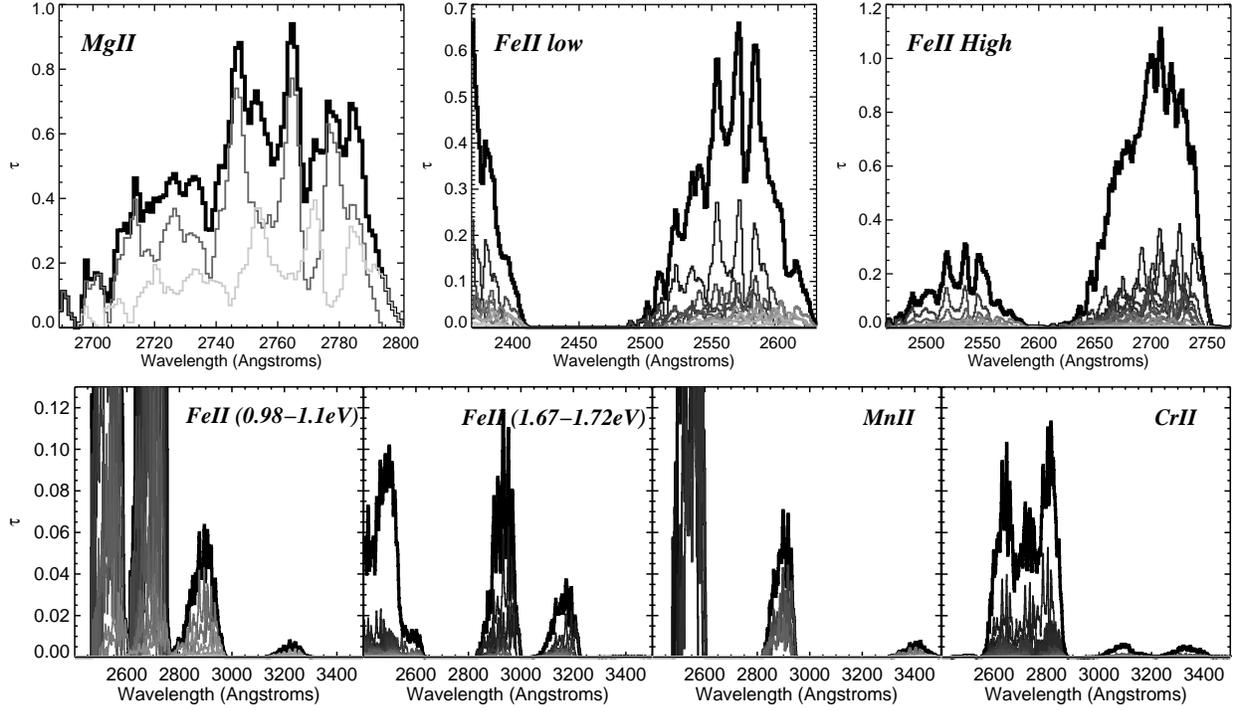}
\caption{\footnotesize Development of the composite absorption profiles; the top panels pertain to profiles used for fitting in \S\ref{abs_modeling}, while the bottom panels pertain to the $\log(N_H)-\log(U) = 23.4$ upper limit in \S\ref{upper_limit}.  In each  panel, the composite profile is
  shown in the   thick black line, while the components making up that
  profile are shown in shades of gray. \label{FIG5} } 
\end{figure}

Then we resampled each $\tau_{j,proto}(v)$ onto a wavelength scale,
with velocity giving blueshift from the rest wavelength of each line
$j$. The net effect was a \ion{Mg}{2} optical depth profile shape
that, as seen (along with \ion{Fe}{2}\,Low and \ion{Fe}{2}\,High)
in the top row of Fig.~\ref{FIG5}, took into account the relative strengths of the
component doublet lines, and could be fit to the spectrum with a
parameter linearly scaling the optical depth $\tau(\lambda)$ .

For \ion{Fe}{2}\,Low and \ion{Fe}{2}\,High, wherein each of the many
lines do not all arise from the same lower level, we defined the
weighting factor $R_i$: the ratio of the level $i$ population to the
population of the lowest-energy level in the set. We used level population ratios $R_i$ from {\it Cloudy} modeling (anticipating our more careful {\it Cloudy}
constraints in \S\ref{cloudy}). The lines within each set arise from a very small range of energy levels, and species in a photoionized gas are in approximate LTE when the density is above their critical densities, so these ratios were approximately equal to the ratios of the levels' degeneracies.

To estimate systematic uncertainties associated with the template shape, we tried each of the five different $\tau_{He}(v)$ derived from FBQS~J1151$+$3822 in \citet{leighly11}, including the saturated (and hence covering fraction dominated) 10830\AA\/ profile and four different estimates of the shallow 3889\AA\/ profile described in that paper. As reported in the following section, this template-related uncertainty is relatively small. The spectral models shown in Fig.~\ref{FIG3} use the MDM-28 \ion{He}{1}*$\lambda3889$ profile described in \citet{leighly11}, because it yields slightly better fits  than the others.

\subsubsection{Spectral Fitting Results\label{fit_results}}

We used the IRAF contributed-program {\tt Specfit} \citep{kriss94} to fit
the spectrum.  This program allows the user to input continua,
Gaussian emission lines (in our case, two \ion{Mg}{2} lines), and profiles of optical depth as a function
of wavelength (in our case, \ion{Fe}{2}\,Low, \ion{Fe}{2}\,High, \ion{Mg}{2}, and for our two strong-iron-emission composite continua, \ion{Mg}{1}). It outputs the $\chi^2$ best fit linear scaling
normalizations for the inputs, along with statistical uncertainties. We performed 15 fits, with each of the three continua and five templates previously mentioned. All normalizations were allowed
to vary, except for that of the PG~1543$+$489 model continuum. Uncertainties in all varying normalizations were propagated into each other.  All features were fixed in wavelength.

A representative spectral fit, using the previously mentioned preferred continuum (\S\ref{other_cont}) and template (\S\ref{profile}), is shown in the top-left panel of Fig.~\ref{FIG3}. The usefulness of splitting \ion{Fe}{2} into the Low and High groups
is clear to see in this fit, where \ion{Fe}{2}\,High is anchored by
the spectrum just short of \ion{Mg}{2}.

We used the output normalizations to extract the net apparent column
densities of the relevant sets of species: Fe$^+$\,Low (0-0.12 eV),
Fe$^+$\,High (0.98-1.1 eV), ground-state Mg$^+$, and ground-state
Mg$^0$. The gist of this calculation is given in Equation 9 of
\citet{ss91}; i.e., for a given energy level $i$, the apparent column
density of level $i$ is
\begin{equation}
N_i=\frac{m_e c}{\pi e^2} \frac{1}{\lambda_j f_j} \int \tau_j(v)
dv,
\label{equation3}
\end{equation}
where $m_e$ is the electron mass, $e$ is the electron charge, $c$ is
the speed of light, and $\tau_j$ is the fitted apparent optical depth
of a single line $j$ from level $i$.  {\tt Specfit} modeling output the optical depth normalization $S$ for each 
absorption profile (\ion{Mg}{2}, \ion{Fe}{2}\,Low,
\ion{Fe}{2}\,High, \ion{Mg}{1}).  We used Equation~\ref{equation1} to define
the fitted optical depth for each of these as  
\begin{equation}
\tau_{j}(v)=S \tau_{j,proto}(v)=S R_i \frac{\lambda_j f_j}{\lambda_{He} f_{He}}
\tau_{He}(v).
\label{equation2}
\end{equation}
Combining Equation \ref{equation3} and Equation \ref{equation2}, and summing over energy levels, the total column density for the level set (the
sum of the column densities for all components of, e.g., \ion{Fe}{2}\,Low), is
\begin{equation}
\displaystyle N=\sum_{i}N_{i}=S \bigg (\sum_{i} R_i \bigg )
\frac{m_e   c}{\pi e^2} \frac{1}{\lambda_{He} f_{He}} \int
\tau_{He}(v) dv.
\label{equation5}
\end{equation}

\begin{deluxetable}{llccc}
\tablecaption{Measured and Modeled Low-Ionization Column Densities}
\tablehead{\colhead{Ion} &
\colhead{Energy} &
\colhead{~} & 
\colhead{Log$_{10}$ Column} &
\colhead{~} \\
& \colhead{} & Apparent\tablenotemark{a} (\S\ref{abs_modeling}) & Lower Limit (\S\ref{fom}) & Upper Limit (\S\ref{upper_limit})}
\startdata
\ion{Fe}{2}\,Low &(0--0.12 eV) & $15.29\pm 0.1$ & 15.38 &  17.64 \\
\ion{Fe}{2}\,High &(0.98-1.1 eV) & $14.89\pm 0.1$ & 14.62 & 16.65 \\ 
\ion{Fe}{2} &(1.67--2.89 eV) & -- & -- & 15.98 \\
\ion{Mg}{2} &(0 eV)\tablenotemark{c} & $15.03\pm 0.1$ & 15.89 & 17.00 \\
\ion{Mg}{1} &(0 eV) & $13.8\pm 0.15$\tablenotemark{b} & 13.65 & 13.65 \\
\ion{Mn}{2} &(0 eV) & -- & -- & 14.85 \\
\ion{Mn}{2} &(1.17 eV) & -- & -- & 13.86 \\ 
\ion{Cr}{2} & (all) & -- & -- & 16.04 \\
\enddata
\tablenotetext{~}{\footnotesize Blanks (--) denote negligible column density, i.e., no effect on the spectrum.}
\tablenotetext{a}{\footnotesize Errors are conservatively altered to be symmetric in log.}
\tablenotetext{b}{\footnotesize \ion{Mg}{1} was not required using the continuum with weakest iron emission, so its detection is tentative. The non-detection is ignored in the stated uncertainty.}
\tablenotetext{c}{\footnotesize Virtually all Mg$^+$ is in the ground state, because its first excited state is $\sim 4$ eV.}
\label{table_ioncolumns}
\end{deluxetable}
These apparent column densities, along estimated errors, are given in the first row of Table~\ref{table_ioncolumns}. The reported column densities are the median results from our 15 fits, and the reported errors delineate the full range from the maximum column plus statistical error to the minimum column minus statistical error (altered in a conservative way to be symmetric in the log, for convenience in {\it Cloudy} modeling). Systematic uncertainty dominates over statistical uncertainty in these errors, but nevertheless is very small. For \ion{He}{1}* in the ensuing physical analysis, we used the average column
density derived in \citet{leighly11}. However, the errors in the
\ion{He}{1}* column density obtained in that paper include the
uncertainty in the covering fraction as well as the uncertainty in the
optical depth; in order for the measurement of the metastable helium
column to have equal weight in {\it Cloudy} model fits, we made the
error on \ion{He}{1}* commensurate with the other ions at $\log
N_{HeI*}=14.9 \pm 0.1$.

The reduced-$\chi^2$ were much larger than one for all fits, a
consequence of including only statistical errors in the modeling when,
in fact, systematic errors dominate. However, the fact that the gross features of the spectrum are fit fairly well, and the fact that there is good agreement
among the results from various fits, suggest that the measured column
densities are of sufficient quality to use as stepping stones to
physical constraints in \S\ref{cloudy}. 

\subsection{Variable Absorption Lines Short of 2960 \AA\/ \label{new_lines}}

\begin{figure}[h]
\epsscale{1.0}
\includegraphics[width=6.5in]{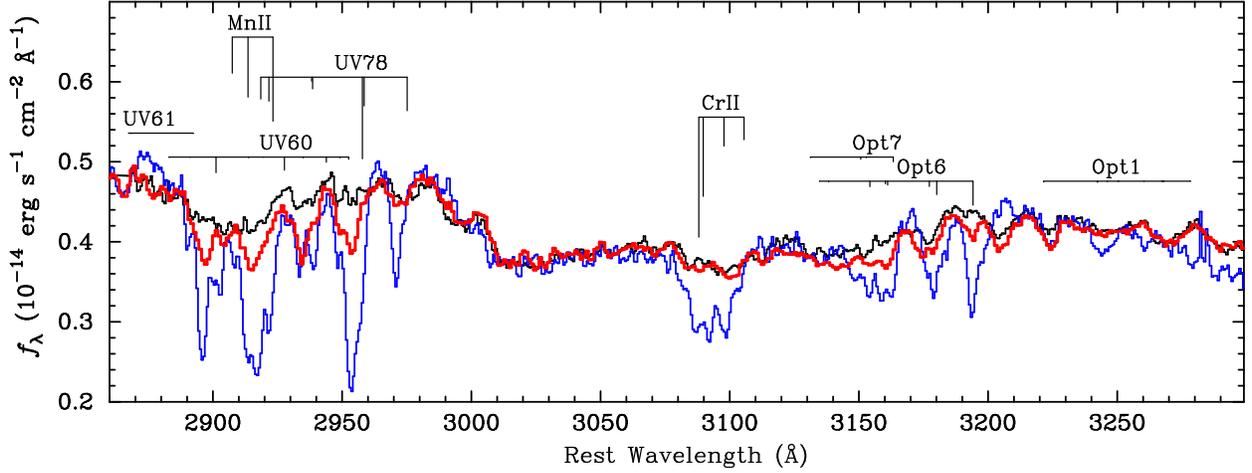}
\caption{\footnotesize New (i.e., variable) narrow absorption lines just short of
  2960\AA\/.  The 2011 KPNO spectrum of FBQS~J1151$+$3822 (red), the 2005 SDSS
spectrum of FBQS~J1151$+$3822 (black), and the SDSS spectrum of 
FBQS~J1214$+$2803 (blue), which has very similar 
(albeit stronger) features.  Lines from \ion{Fe}{2} (labeled with their multiplet designations from \citet{moore52}), \ion{Mn}{2} and \ion{Cr}{2}, shifted by 
$-3,300\rm \, km\, s^{-1}$ are marked, where the tick length is proportional to the 
oscillator strength for each ion. The higher excitation of some of the new multiplets requires a high density, suggesting the condensation of dense cores in the outflow as discussed in \S\ref{plausible}.  \label{FIG6} }
\end{figure}

The SDSS spectrum, taken on 2005 March 3, samples a wavelength range
extending down to $\sim 2850$\AA\/ in the rest frame.  We show the near UV portion of the new (2011 May) KPNO spectrum overlaid on
the SDSS spectrum in
Fig.~\ref{FIG6}.  There is a good correspondence overall, but 
the new spectrum has additional absorption lines just short of 2960\AA\/,
and possibly also near 3200\AA\/.  We used spectral fitting to measure
the centroid positions of the four features near 2960\AA\/: 2896,
2916, 2933, and 2952\AA\/.   

These lines were not trivial to identify without additional
information.  Fortunately, FBQS~J1214$+$2803 \citep[analyzed by][]{dekool02a} has very
similar features, slightly shifted in wavelength, and we overlay the
SDSS spectrum of that object in Fig.~\ref{FIG6}.  Following the line
identifications given by \citet{dekool02a}, we found that the features
near 2960\AA\/ are a combination of excited \ion{Fe}{2} from 
transitions with lower levels near 1~and~1.7~eV, and excited state
\ion{Mn}{2} with lower levels near 1.2~eV, with a blueshift of around
$3400\rm \, km\, s^{-1}$. 

As discussed in more detail in \S\ref{upper_limit}, these new lines have very low opacity, a consequence of small oscillator strength, sparsely populated excited states, and/or low abundance (depending on the line in question). Variability apparently limited to low-opacity transitions seems to be unprecedented in the literature. However, we have only one epoch shortward of $\sim
2850$\AA\/.  It may be that high-opacity lines like \ion{Mg}{2} and \ion{Fe}{2}
shortward of 2850\AA\/ also vary.  

At first glance, and given the velocity
width of the \ion{He}{1}* lines (e.g., Fig.~\ref{fig2}), it is not
obvious why relatively narrow lines would appear at the observed
velocity offset.  However, the \ion{He}{1}* profile can plausibly be
broken into two parts, a low-velocity component between $\sim -3,000$
and $-500\rm \, km\, s^{-1}$, and a high-velocity component between
$\sim -11,000$ and $\sim -3,000\rm\, km\, s^{-1}$.  So these narrow
features are located at the low-velocity end of the higher-velocity
feature.  That type of structure---thicker, lower-ionization gas lying
at lowest velocities for a given component---appears to be common in
BALQSOs \citep{voit93}.  Furthermore, examination of the
$dN_{HeI*}$ as a function of velocity for the partial covering model
of \citet[][Fig.\ 6]{leighly11} shows a suggestive maximum near
$-3,300\rm \, km\, s^{-1}$ that may support higher column density at
this velocity.

\section{Physical Analysis\label{cloudy}}

In \citet{leighly11}, we performed {\it Cloudy}
\citep{ferland98,ferland13} photoionization simulations to constrain the physical
conditions in the gas responsible for the \ion{He}{1}* BAL in
FBQS~J1151+3822, including hydrogen density $n$ (atoms per $cm^3$), ionization parameter $U$, and hydrogen column density $N_{H}$ (per
$cm^2$) --- tending to use $\log N_{H}-\log U$ in place of this last parameter, so that the hydrogen ionization front lies at a constant value. We found that the ionization parameter had to be
higher than $\log U \sim -1.4$ to produce
sufficient \ion{He}{1}*. Higher-resolution {\it
  Cloudy} runs have since reduced that minimum to $\sim -1.5$. The number density was constrained to be less
than $n\sim 10^8\rm\, cm^{-3}$ by the absence of Balmer absorption
lines in our spectra. We could not further constrain the gas
parameters directly, although we made a dynamical argument for a
minimum density of $\log n \sim 7$ under the assumption that the outflow was accelerated by radiative
line driving. 

In this section, using the same {\it Cloudy} setup, we describe further constraints on the gas properties
imposed by the low-ionization absorbers in our new KPNO spectrum; these allowed us to lift the assumption of acceleration by radiative line driving and still obtain useful constraints.

We did, however, assume step-function/homogeneous covering, except where otherwise noted, so that the ratio of
the observed spectrum to the continuum in a single absorption line is given by
\begin{equation}
\frac{I}{I_0}=\exp({-\tau_{app}})=1-C_f+C_f \exp(-\tau_{true})
\label{equation6}
\end{equation}
\citep{hamann97,arav05}, where $\tau_{app}$ is the apparent
optical depth considered in \S\ref{analysis}, $C_f$ is the covering
fraction of the absorber, and $\tau_{true}$ is the
true optical depth. In this model, the apparent columns measured in \S\ref{analysis} are a lower limit for the average column density (the product of true column and covering fraction). As is standard practice \citep[e.g.,][]{sabra05}, we took ionic columns in our photoionization models to refer to this average column, as an implicit compromise between innumerable possible spatial geometries. We explore inhomogeneity in \S\ref{plausible}.

We also assumed {\it Cloudy}'s default solar abundances, as given in Hazy, the {\it Cloudy} manual.

\subsection{The Utility of the \ion{He}{1}*/\ion{Fe}{2} Combination \label{prelim_cloudy}}

\begin{figure}[h]
\epsscale{1.0}
\includegraphics[width=6.5in]{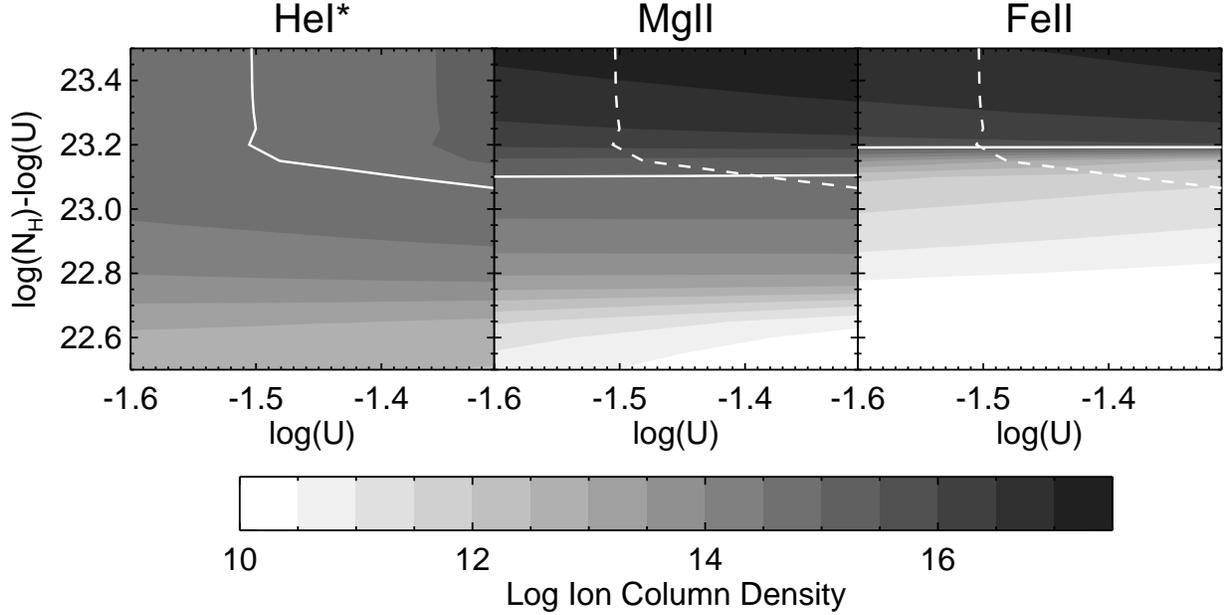}
\caption{\footnotesize Investigation of {\it Cloudy} parameter space, discussed in
  \S\ref{prelim_cloudy}.  Each panel shows   shaded contours of ion
  column density. The white solid contours show our measured average
  (\ion{He}{1}*) and apparent (\ion{Mg}{2}, \ion{Fe}{2}) column
  densities of the various ions, where \ion{Fe}{2} is 
  the sum of \ion{Fe}{2}\,Low and \ion{Fe}{2}\,High.  The
  dashed white contours replicate the \ion{He}{1}* measured average
  column contour for 
  comparison. The intersection of the \ion{He}{1}* and \ion{Fe}{2} white contours show that $\log U \sim -1.5$ and that $\log N_H - \log U$ is at least $\sim -23.2$. The \ion{Mg}{2} white contour is below this intersection, indicating \ion{Mg}{2} saturation. \label{FIG7}}
\end{figure}

Fig.~\ref{FIG7}
shows contours of average ion columns\footnote{\footnotesize A similar plot appears
  in the proceedings paper \citet{leighly12}.} with superimposed
measurements for these absorbers, as a function of ionization
parameter and $\log N_H - \log U$ for a constant density of $\log n=
7$.  The measured \ion{He}{1}* average column becomes approximately
vertical at $\log U \sim -1.5$; here, the \ion{He}{1}* column density
is ionization bounded near the hydrogen ionization front. That is, for a
typical AGN continuum, no He$^+$ ions exist beyond the hydrogen
ionization front, and since \ion{He}{1}* is created by recombination
onto He$^+$ (the ratio of metastable He$^0$ to He$^+$ is nearly constant), no \ion{He}{1}* can exist there either. For normal
quasar SEDs, He$^+$ dominates the ionization balance in the outer
portion of the \ion{H}{2} region \citep[e.g., Fig.\ 1 in][]{hamann02}.
Like hydrogen, for a semi-infinite slab, the amount of He$^+$ depends
nearly linearly on the ionization parameter.  The more
intense the photon field, the larger the \ion{H}{2} region, the more ionized helium ions are present, and the
more metastable helium ions are present.  So the  measured 
\ion{He}{1}* column density sets a lower limit on the ionization
parameter. 

Hence, at the minimum $\log U
\sim -1.5$, \ion{He}{1}* attains the measured average column at the
hydrogen ionization front, and increasing the column density beyond
the front cannot increase the \ion{He}{1}*.  This is not true for
ionization parameters $>-1.5$; there, the gas must be matter bounded (truncated somewhere before the hydrogen ionization front)
in order to not exceed the measured \ion{He}{1}* average column
density. 

The \ion{Mg}{2} and \ion{Fe}{2} column density contours are approximately parallel to the
hydrogen ionization front; to produce these low-ionization lines with
a high ionization parameter, there simply needs to be a commensurately
high hydrogen column. \ion{Fe}{2} is very rare in the gas for
lower values of $\log N_H$ at a given $\log U$, becoming significant
only by the vicinity of the hydrogen ionization front. 

These properties imply that the combination of \ion{He}{1}* plus \ion{Fe}{2}
together provide unparalleled constraints on both column density and
ionization parameter. Some of the diagnostic powers of this
combination have previously been noted by, e.g.,
\citet{korista08}. Here, however, we emphasize that the ions occur in disjoint
regions of the gas; \ion{He}{1}* requires He$^+$ to be present, and
that is only found in the \ion{H}{2} region, while \ion{Fe}{2}
requires penetration of the hydrogen ionization front; the presence of \ion{Fe}{2} implies that \ion{He}{1}* is ionization bounded. So where both
\ion{He}{1}* and \ion{Fe}{2} are present, as they are in our object, 
the former's column yields the specific ionization parameter, and the latter's
column constrains the hydrogen column. 

Given these tools, even a cursory glance at the \ion{He}{1}* and
\ion{Fe}{2} panels of Fig.~\ref{FIG7} will correctly suggest that $\log U \sim
-1.5$ and that $\log N_H - \log U$ is at least $\sim -23.2$.  Then, in
light of these requirements, the \ion{Mg}{2} panel implies that the
measured \ion{Mg}{2} apparent column is definitely less than its
average column by a factor large enough to imply that it is
saturated; in contrast, the \ion{Fe}{2} apparent column is not
obviously inconsistent with the model.

\subsection{Figure-of-Merit Fit from the Measured Columns\label{fom}}

Following up on these suppositions, our next step was to find the gas
parameters that best fit the measured columns,
using a figure-of-merit analysis.  For our figure of merit, we used the
absolute value of the difference between the measured ion column
densities and the {\it Cloudy} model column densities, divided by the
measurement uncertainty in the column density.  We had used a similar
figure of merit previously \citep{casebeer06, leighly04}, and found that
it handles outliers better than $\chi^2$, although the statistical
significance of the value is not as clear.  We considered the average
column density of \ion{He}{1}*, the apparent column densities of
\ion{Fe}{2} Low, \ion{Fe}{2} High, and \ion{Mg}{1}, and the upper
limit on Balmer absorption obtained in \citet{leighly11}. Since
\ion{Mg}{2} is unambiguously saturated, we excluded its apparent
column from the fit.  

\begin{figure}[p]
\epsscale{1.0}
\includegraphics[width=4.0in]{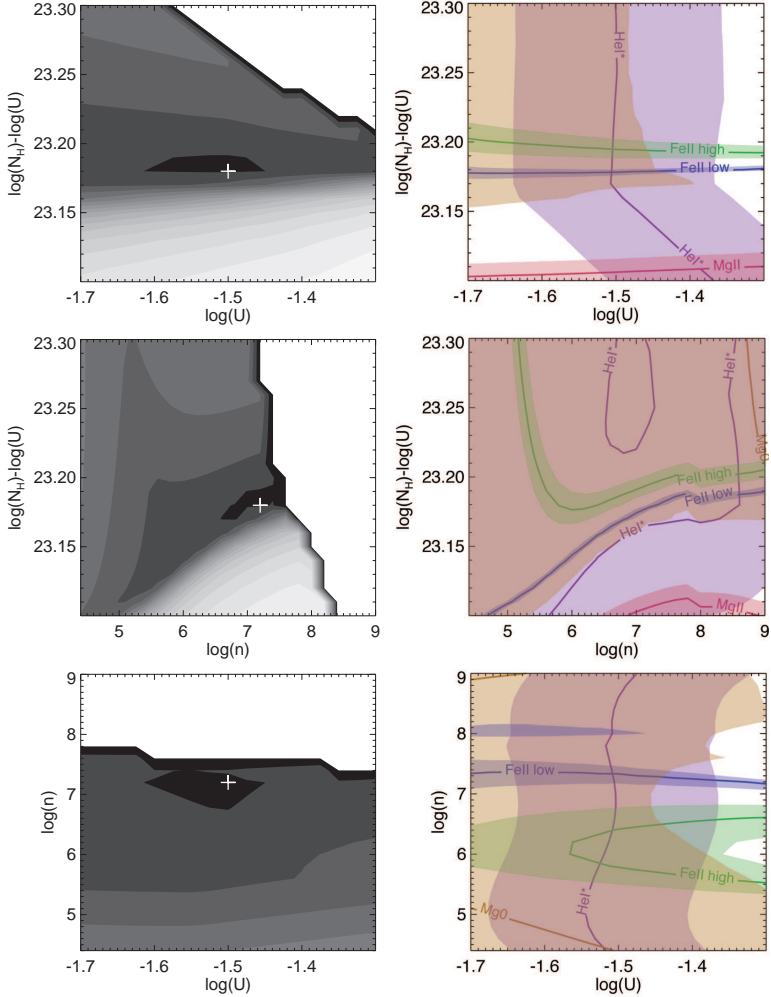}
\caption{\footnotesize The results of the figure-of-merit analysis, showing
  contours through cross sections of parameter space centered around
  figure-of-merit minimum marked by the plus sign and located at $\log 
  U=-1.5$, $\log n=7.2$, and $\log N_H-\log U = 23.18$, corresponding
  to $\log N_H=21.68$.   The left side
  shows contours of the figure of merit.  The white regions show where
  the  predicted Balmer absorption is larger than the upper limit
  implied   by non-detection of H$\alpha$ absorption \citep{leighly11}, and are therefore ruled
  out.  The strong gradient of the contours bordering  that region are
  an artifact of the plotting program.   The right side shows the contours of the
  column densities of the measured ions.  The solid line shows the
  measured value, and the shaded areas show the uncertainty range.
  The contours are colored as follows: \ion{He}{1}* is purple,
  \ion{Fe}{2} Low is blue, \ion{Fe}{2} High is green, \ion{Mg}{2} is
  red, and \ion{Mg}{1} is taupe. As discussed in \S\ref{fom}, the \ion{Fe}{2}\,High and Low contours on the right side suggest useful constraints on $\log n$, while the non-intersection of these two contours hints that \ion{Fe}{2} may be saturated. \label{FIG8}} 
\end{figure}

The results are shown in Fig.~\ref{FIG8}.  The left side of
Fig.~\ref{FIG8} shows contours of the figure of merit.  The + sign
marks the position  of merit minimum, located at $\log
U=-1.5$, $\log n=7.2$, and $\log N_H-\log U = 23.18$ (corresponding to
$\log N_H=21.68$); the middle row of Table~\ref{table_ioncolumns} lists the predicted ion columns at this best fit. The white areas are regions in which the Balmer
absorption upper limit would be violated. As expected, column
densities below the best fit are strongly excluded due to the \ion{Fe}{2} behavior discussed in the preceding section.

The right side of Fig.~\ref{FIG8} shows contours of ionic column
density as a function of the gas properties.  The solid contours show
where the observed ionic column density equals the model ionic column
density, while the shaded regions of the same color show the
measurement uncertainty ranges.  

The middle-right panel shows that the data impose interesting
constraints on the number density.  As noted by, e.g.,
\citet{korista08}, excited state \ion{Fe}{2} lines have critical
densities around $\log n \sim 4.5$, while ground-state iron does not
have this restriction.  This means that very little \ion{Fe}{2}\,High
is present in the gas for log densities lower than $\sim 5$.  But for
log densities higher than $\sim 6$, the \ion{Fe}{2} Low and
\ion{Fe}{2} High contours begin to approach one another, then run
nearly parallel to each other through high densities, up to and beyond the limit imposed
by the lack of Balmer absorption. The top-right panel, likewise, illustrates the
robust constraints on ionization parameter imposed by \ion{He}{1}*. 

The right panels also constitute our first intimation that \ion{Fe}{2} is saturated; if the \ion{Fe}{2} apparent columns were approximately equal to the average columns, \ion{Fe}{2} High and
\ion{Fe}{2} Low contours or their error bars would
overlap.  They do not.

We created $C_f=1$ spectral models using the ionic columns from {\it
  Cloudy} at the figure-of-merit minimum, together with the templates
and continua discussed in \S\ref{abs_modeling}. That is, inverting the
procedure described in \S\ref{fit_results}, we took the
{\it Cloudy} ion columns and calculated the corresponding scaling
factors for the optical depth profiles. These models, exemplified in
the upper-right panel of Fig.~\ref{FIG3}, demonstrated that
\ion{Mg}{2} is completely saturated, and reaffirmed that the {\it Cloudy} figure-of-merit best fit
predicts insufficient \ion{Fe}{2}\,{High}.

\subsection{\ion{Fe}{2} Saturation Near the H Ionization Front \label{saturation}}

As discussed above, our best-fitting {\it Cloudy} model predicts too
little \ion{Fe}{2}\,High relative to \ion{Fe}{2}\,Low.  To investigate
this further, we searched parameter space for physical conditions that
would produce columns of \ion{Fe}{2} closer to the measured
apparent columns. The measured columns from the phenomenological fit (\S\ref{fit_results})
predict a ratio of 2.5 for the columns of \ion{Fe}{2}\,Low /
\ion{Fe}{2}\,High.  However, the figure-of-merit best fit in {\it Cloudy} parameter space
(see \S\ref{fom}) yielded a ratio of 5.6. Broadening the search to minimize the figure-of-merit
over {\it only} \ion{Fe}{2} species (i.e.,
$|FeIILow(obs)-FeIILow(Cloudy)|+|FeIIHigh(obs)-FeIIHigh(Cloudy)|$), we
found a ratio of 5.0, still a factor of two higher than the ratio
measured and barely better than 5.6. Thus, anywhere in
parameter space that would produce the right amount of
\ion{Fe}{2}\,Low column density, too little \ion{Fe}{2}\,High is
predicted.

\begin{figure}[h]
\epsscale{1.0}
\includegraphics[width=6.5in]{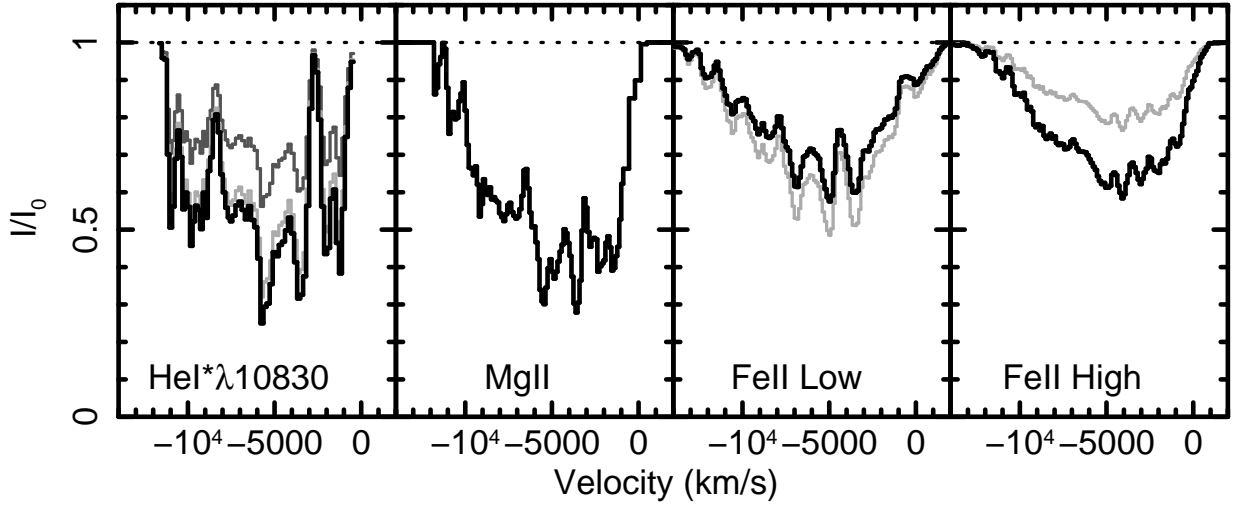}
\caption{\footnotesize The {\it modeled} ratio $I/I_0$ for various absorption lines. {\it Far
    Left:} The \ion{He}{1}*$\lambda 10830$ ratio.  The dark gray line
  shows the original profile presented in \citet{leighly11}. The light
gray and black lines show the profile after subtracting the galaxy and
the galaxy plus torus (Appendix~\ref{host_gal}), respectively. {\it Middle Left:} The \ion{Mg}{2} model ratio
from the phenomenological fit model presented in \S\ref{fit_results}.   The velocity is computed using the
oscillator-strength-weighted mean wavelength for the doublet. {\it
  Right Two Panels:} The \ion{Fe}{2} model ratios.  The black lines show
the ratio from the phenomenological fit model presented in \S\ref{fit_results}.  The gray lines show the levels predicted
for the figure-of-merit best fit for the {\it Cloudy} modeling presented in \S\ref{fom}. As discussed in \S\ref{saturation}, the right two panels suggest that \ion{Fe}{2}\,Low and High are saturated. \label{FIG9}}
\end{figure}

To explore this discrepancy further, we show in Fig.~\ref{FIG9} the ratios of model-data to the continuum for
\ion{He}{1}*$\lambda 10830$, \ion{Mg}{2}, \ion{Fe}{2}\,Low, and
\ion{Fe}{2}\,High. Best fit results from spectral fitting, and model
results from {\it Cloudy}, are shown. Despite their different columns
and line strengths, the \ion{He}{1}*$\lambda 10830$ and \ion{Mg}{2}
fits are about as deep as one another, a consequence of the lines being
saturated and having roughly similar (though not identical---see
\S\ref{upper_limit} for \ion{Mg}{2} and Appendix \ref{host_gal} for \ion{He}{1}*) covering fractions. Likewise, the two
\ion{Fe}{2} complexes also reach approximately the same depth as each
other, albeit a much lower depth than the \ion{Mg}{2} and \ion{He}{1}*
lines. This suggests that their apparent optical depths are influenced much more
by covering fraction than by the optical depth of the gas, and that \ion{Fe}{2}\,Low, and likely
\ion{Fe}{2}\,High as well, are saturated.

This result does not appear to be an artifact of the modeling.  
Differences caused by using other continua are not significant, and nor are our conclusions a consequence of blending with \ion{Mg}{2},
since the high-velocity side of \ion{Fe}{2} High is unblended (see
Fig.~\ref{FIG3}, upper-right panel).

\begin{figure}[h]
\epsscale{1.0}
\includegraphics[width=6.5in]{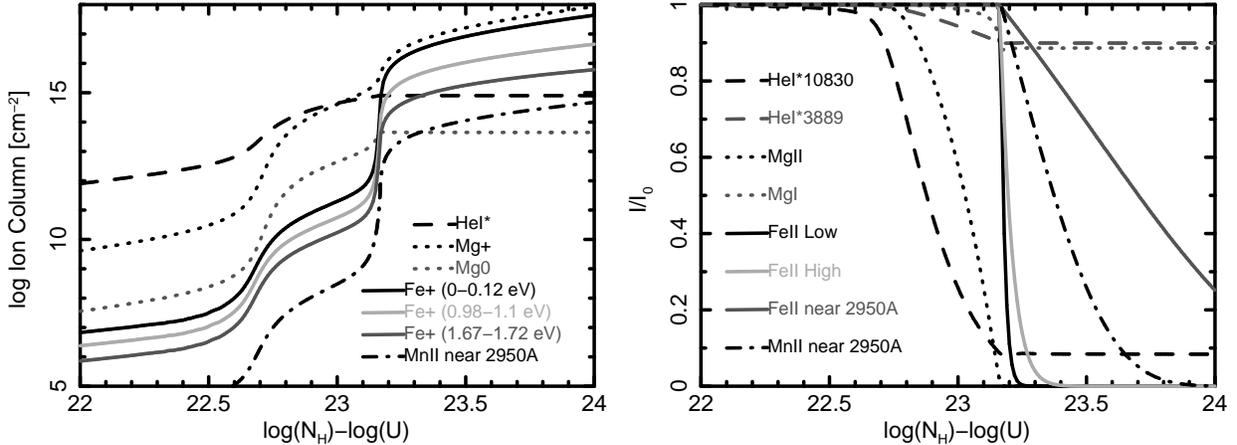}
\caption{\footnotesize Ion columns and resultant $I/I_0$ ratios as a function of depth into the absorbing gas, from {\it Cloudy} modeling. {\it Left:} Predicted log integrated ion column density for $\log U=-1.5$ and $\log n=7.2$ as a function of $\log N_H - \log U$. {\it Right:} The same, but with the ion column replaced by the resultant $I/I_0$ produced by each ion in the line complexes of interest.  The absorber is assumed to fully
  cover the source, for a rectangular absorption profile with a
  width of $5,000\rm \, km\, s^{-1}$.  These lines show at what depth
  in the gas the lines will saturate. Among other things, these panels demonstrate the general likelihood of \ion{Fe}{2} saturation, the physically distinct regions over which \ion{He}{1}* and \ion{Fe}{2} produce opacity, and the physical distinction between HiBALs, LoBALs, and FeLoBALs as a matter of gas thickness relative to the hydrogen ionization front. The hydrogen ionization front occurs just short of $\log N_H - \log U = 23.2$. \label{FIG10}}
\end{figure}

Indirect support for \ion{Fe}{2} Low and \ion{Fe}{2} High saturation
comes from the right panel of Fig.~\ref{FIG10}, which
shows predicted $I/I_0$ for various absorption complexes as a function
of $\log N_H-\log U$, for our best fit values of $\log U=-1.5$ and
$\log n=7.2$.  To construct this figure, we assume a rectangular
absorption profile with a velocity width of $5000\rm \, km\, s^{-1}$;
this is loosely equivalent to the profiles we observe, with velocity
widths of $\sim 11,000\rm \, km\, s^{-1}$ and covering fractions of
$\sim$50\%.  $I/I_0$  features comprised of multiple lines were
treated as a single line characterized by the sum of the $\tau$ values
for each line; this is an approximation, since opacity from a group
of lines will be spread over a larger wavelength than a single line, but plot is meant to be qualitative.

In this plot, $I/I_0$ values for the \ion{Fe}{2} complexes drop, and then level off to saturation, very
rapidly at the hydrogen ionization front.  This suggests that the spectra of a hypothetical sample of outflows spanning a
range of $\log N_H-\log U$ would, if they had sufficient number density to produce \ion{Fe}{2}\,High, generally either have no \ion{Fe}{2}
absorption, or have saturated \ion{Fe}{2}\,Low and
\ion{Fe}{2}\,High. 

In any case, it is clear that the Fe$^+$ apparent columns,
like the Mg$^+$ apparent column, are only lower limits on their
respective average columns. Likewise, the best fitting hydrogen column
from these apparent columns is only a (very rigorous) lower limit.

An interesting result for BALQSO classifications also emerges from these plots, which suggest that the difference between FeLoBALs, \ion{Mg}{2} LoBALs, and HiBALs  could be simply column density relative to the column density of the \ion{H}{2} zone.  Specifically, if \ion{Mg}{2} is present
and \ion{Fe}{2} is absent, the outflow column density is truncated
before the hydrogen ionization front is breached, while the presence
of \ion{Fe}{2} requires penetration of the hydrogen ionization front. If all low-ionization lines are absent, the outflow column density is truncated before \ion{Mg}{2} can populate. 

This behavior has a physical explanation. Because photons of only $7.6\rm \, eV$ and $15\rm \, eV$ are sufficient to form, respectively, Mg$^+$ from Mg$^0$ and Mg$^{+2}$ from Mg$^+$, it is
often assumed that Mg$^+$ accumulates a
significant column density only at depths beyond the hydrogen
ionization front, disjoint from the HiBAL-forming gas. However,  the $80\rm \, eV$ photon required to form Mg$^{+3}$ from Mg$^{+2}$ means that Mg$^{+2}$ dominates at intermediate depths before the hydrogen ionization front and after the depletion of 80\rm \, eV photons, a HiBAL-producing region which includes the He$^+$ zone. Recombination onto this Mg$^{+2}$ inevitably produces some Mg$^+$, and because of the high abundance of magnesium (plus the high oscillator strengths of the 2800 \AA\/ doublet), this is enough to make an observable, or even strong, line.

The first two ionization potentials of iron (7.9 and 16.2 eV) are quite similar to their magnesium counterparts, but only 30.65 eV is required to form Fe$^{+3}$ (and 54.8 eV to form Fe$^{+4}$, similar to the 54.4 eV to form  He$^{+2}$). Thus, in the vicinity of the He$^+$ zone, where Mg$^{+2}$ dominates over other ionization states, iron is distributed widely between Fe$^{+2}$, Fe$^{+3}$, and perhaps some marginal Fe$^{+4}$, so that much less Fe$^+$ forms from recombination. The Fe$^+$ column is also distributed amid a greater number of low-energy excitation states than is the Mg$^+$ column, with many transitions from those levels, so that observable opacity is even more difficult to develop.

\subsection{Obtaining an Upper Limit on the Column Density\label{upper_limit}}

Saturation of \ion{Fe}{2}\,Low and \ion{Fe}{2}\,High
(\S\ref{saturation}) means that the best-fitting hydrogen column
derived in \S\ref{fom} is only a lower limit. But, as discussed in
\S\ref{new_lines}, there are several relatively shallow \ion{Fe}{2} lines just short of
2960\AA\/, corresponding to transitions from the 0.98--1.12 and
1.67--1.71~eV excitation states of \ion{Fe}{2}. The 1.7~eV levels are
highly excited and thus are expected to have low populations, but their lines have higher
oscillator strengths than those of the 1~eV lines at these wavelengths (in our final model, the ratio of 1.7 to 1 eV contribution---measured by integrated $\tau_{app}$ from 2852 to 3000 \AA\/---is 1.6 to 1). In addition, there are three important
\ion{Mn}{2} transitions from excited state levels at 1.17 eV. Manganese is an iron-peak element about 100
times less abundant than iron, but these transitions have high oscillator strengths and
moderate excitation (the ratio of \ion{Mn}{2} to 1 eV \ion{Fe}{2} contribution is 0.94 to 1). Assuming for the moment that these 2960\AA\/ lines have the
same covering fraction as their \ion{Fe}{2} counterparts short of
2800\AA\/, they are sufficiently shallow relative to the covering
fraction so as to provide information on the average optical depth,
and thus lead to upper limits on column densities. This comes from the fact, demonstrated in \citet{leighly11}, that for lines absorbing much less than the covering fraction, the apparent optical depth is approximately equal to the average optical depth.

At the hydrogen columns necessary to produce these lines, \ion{Cr}{2} and highly excited \ion{Fe}{2} also reach significant opacities
in various sections of the UV spectrum, and are included in our
model. We included all \ion{Fe}{2} lines in the bandpass with
oscillator strength and air wavelength data from NIST, with lower
levels up to 3 eV. \ion{Mn}{2} data were likewise taken from NIST,
except for a strong ground-state line at 2576\AA\/; this line was
missing from NIST, so we included its data from the Kurucz Atomic Line
Database\footnote{\footnotesize http://www.cfa.harvard.edu/amp/ampdata/kurucz23/sekur.html}. NIST
information on \ion{Cr}{2} is very incomplete, so we took data for
this ion from the Kurucz database, including all lines from $<$ 3 eV
with $\log(gf) > -3.0$.

Although {\it Cloudy} does not output the level-by-level column densities for \ion{Mn}{2} and \ion{Cr}{2} as it does for \ion{Fe}{2}, both Manganese and Chromium are iron-peak elements, so at a given position, the populations of their excited states can be assumed to have the same multiplicative departure from LTE as \ion{Fe}{2} does. Thus, we obtained \ion{Cr}{2} and \ion{Mn}{2} departure coefficients, for each energy level and for each position along the line-of-sight length of the gas, by linearly interpolating over energy between the \ion{Fe}{2} departure coefficients of the two nearest \ion{Fe}{2} energy levels. Together with the ion fractions and temperatures provided by {\it Cloudy} incrementally along the length of the gas, we used these to calculate the net column densities of energy states as the integrals over length of the product of departure coefficient and LTE density.

\begin{figure}[h]
\epsscale{1.0}
\includegraphics[width=3.5in]{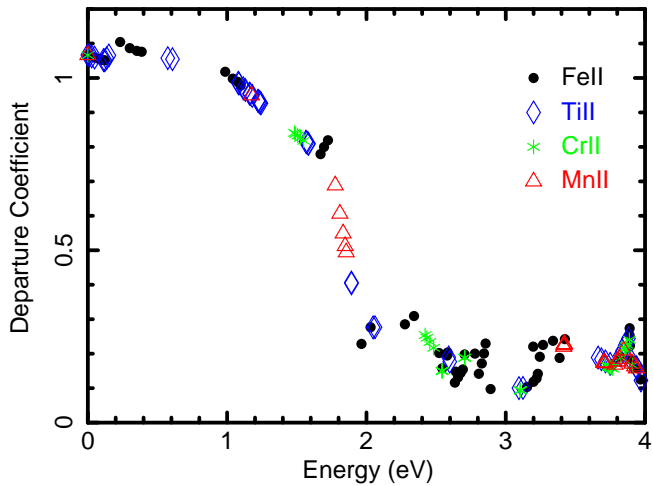}
\caption{\footnotesize Aggregate departure coefficient (i.e., our integrated non-LTE column density divided by our integrated LTE column density) as a function of energy level above ground, for several once-ionized iron-peak elements. Depletion by non-LTE effects rapidly becomes insignificant below $\sim 2$~eV. \label{FIG11} } 
\end{figure}

We computed these ion columns for a range of hydrogen columns. Levels from excited states greater
than $\sim 2$~eV above ground are extremely depleted by departure from
LTE, but this effect has minimal impact on the
partition function due to the high energies at which it becomes
significant. So for the $\sim 2960$\AA\/ lines (which are all from
levels below 2 eV), the columns densities are near LTE (Fig.~\ref{FIG11}). We investigated density dependence and found it to be insignificant. We also tested \ion{Ti}{2}, but this ion turned out to have virtually no opacity in our bandpass, so we ignored it. 

Using the continua and \ion{He}{1}* templates described in
\S\ref{analysis}, we constructed the predicted spectrum for a range of
$\log(N_H)-\log(U)$ at $\log n = 7.2$ and $\log U =
-1.5$.  Average column densities for each ion and energy level,
obtained from {\it Cloudy} parameter space as described above, were
translated into average optical depths for each corresponding line
using Equation~\ref{equation3}, which were converted to true optical
depths using $\tau_{avg}=C_f\,\tau_{true}$. These true optical depths
were then converted to into the observed $I/I_0$ using
Equation~\ref{equation6}. We used a covering fraction $C_f$ of 0.25,
corresponding to the average fraction (weighted by the inverse of the observed flux's Poisson variance) of the continuum absorbed between 2500 and 2620 \AA\/, the range over which \ion{Fe}{2}\,Low is {\it completely} saturated at high hydrogen columns.

By the time $\log(N_H)-\log(U)$ reaches 23.4, the model absorption
is deeper than the spectrum by a substantial amount (corresponding to at
least $\sim$ twice the typical systematic errors implied by the few
wavelengths where our continua models drop below the spectrum) for the
unabsorbed data between $\sim 2852$ and $\sim 2890$\AA\/, as well as
for a substantial fraction of the longward absorption features. In fact, at this column density, the $\sim 2852$ to $\sim 2890$\AA\/ criterion was very nearly met for covering fractions as low as $\sim 0.1$. A model spectrum for this column is shown in the lower-left panel of
Fig.~\ref{FIG3}, which was made using a velocity-constant $C_f = 0.25$
for iron-peak absorbers. We used $C_f = 0.58$ for magnesium absorbers, the weighted average fraction of the continuum absorbed between 2690 and 2788 \AA\/, the range over which \ion{Mg}{2} is completely saturated. The ion column densities for this upper limit are tabulated in the last row of Table~\ref{table_ioncolumns}. The contributions of each line to the composite
{\ion{Fe}{2} (0.98--1.12), \ion{Fe}{2}
($1.67$--$2.89\rm \, eV$), \ion{Mn}{2}, and \ion{Cr}{2} average
optical depths at this hydrogen column are shown in the bottom panels of
Fig.~\ref{FIG5}. 

This upper limit model also exemplifies the manner in which saturation
and partial covering can equalize the \ion{Fe}{2}\,Low and
\ion{Fe}{2}\,High absorption complexes, as well as yield the observed profile
of \ion{Mg}{2}. It shows that the covering fraction for \ion{Mg}{2} is more than
twice that of the broad \ion{Fe}{2} complexes, a subject which we 
discuss further in \S\ref{inhomogeneous}. That said, this model is
far from being a complete physical picture; for example, it yields too much absorption from \ion{Mn}{2} and the upper levels of \ion{Fe}{2} at short wavelengths, fits the 2960\AA\/ lines poorly, and yields too much absorption between 2852 and 2890\AA\/. Therefore, we explore inhomogeneous models in the next section.

\subsection{A Possible Model of Inhomogeneity\label{plausible}}

Despite the fact that our models reproduce the spectra
reasonably well, previously-discussed regions of poor fit in the lower (\S\ref{fom}, \S\ref{saturation}) and upper (\S\ref{upper_limit}) limit spectral models, which remain problematic in models with intermediate hydrogen columns, imply that the real situation may be more complicated. Our assumption that the absorption profiles of once-ionized iron-peak elements and \ion{Mg}{2} are the same as
that for \ion{He}{1}* probably contributes to the inconsistencies, as does our assumption of a velocity-constant covering fraction.  In addition, the absorber could be inhomogeneous, so that the apparent step-function covering fraction (i.e., that obtained by assuming homogeneity) will be a convolution of the line strength and the spatial dependence of gas column.

The plausible large number of degrees of freedom,
significant line blending, and poor signal-to-noise ratio mean that no
unique solution is possible.  However, as shown in this section,
further examination of the data can lead us to a possibly plausible,
albeit ad hoc, phenomenological, and unconstrainable, physical picture composed of three outflow components. Our purpose here is not to present a final model, but to propose a possible physical picture and determine the specific weak points of our homogeneous-model kinematic constraints.

We broke the
template profile into three velocity pieces with different opacities
and covering fractions.  First, an intermediate-velocity narrow Gaussian
centered near $\rm -3,300~km\,s^{-1}$ is suggested by the
high-excitation  \ion{Fe}{2} and \ion{Mn}{2} near 2960\AA\/, and probably
has high opacity (to produce these underpopulated lines) and low covering fraction
(since these lines are not deep).  Second, a low-velocity component between
$\sim -2,300$ and $-1,000\rm \, km\, s^{-1}$ is required to model the
deepest part of ground-state \ion{Fe}{2} and \ion{Mg}{2}; this
component cannot have high opacity (because it would produced too much
absorption from 2631\AA\/ \ion{Fe}{2}, an objectionable property of
the upper limit as seen in Fig.~\ref{FIG3}) but must have high covering fraction (since \ion{Mg}{2} is deep).  The final, high
velocity component, between $-11,000$ and $-4,000\, \rm km\, s^{-1}$,
is adapted from the \ion{He}{1}$\lambda 10830$ profile; this
component should have intermediate optical depth and intermediate
covering fraction, as it represents the bulk of the template
fit.  That is, the modifications that we are making in this analysis
influence most heavily the lower velocity components.

To model the composite absorption, we used an inhomogeneous
covering model which models the optical depth $\tau$ as a power law
\citep{arav05,sabra05}, parameterized by $\tau_{max}$ and
power law index $a$ because, unlike the step-function partial
covering, the composite $I/I_0$ is a product of the $I/I_0$ for the
three components.  We varied the power law slope and maximum opacity
by hand, and applied the resulting optical depth profile to the
continuum.   The resulting model is shown in the lower-right panel of
Fig.~\ref{FIG3}.  The parameters are only loosely constrained by
the data, so this is not a formal fit, but rather a plausible
model. However, some illuminating trends emerge.  First, as expected,
for the $-3,200\rm \, km\, s^{-1}$  component, $\tau_{max}$ has to be
large to reach the substantial optical depths required to produce the
lines near 2960\AA\/, but $a$ also has to be large to yield a
sufficiently low covering fraction so that the lines near 2600
and 2700\AA\/ are not too deep. Second, the low-velocity component has to have the smallest value of $a$ to yield a
large enough covering fraction to produce the deep low-velocity
component of \ion{Mg}{2}.  Finally, the high-velocity component has to
have low-intermediate values of $a$ and $\tau_{max}$ to fill in the
remaining \ion{Mg}{2}.  

This inhomogeneous model explains the 2960\AA\/ features
better than any other model we devised. And since the very high  
opacity is limited to the $-3,200\rm\, km\, s^{-1}$ component, it does
not produce too much opacity between 2600 and 2631\AA\/.  That said, this model
produces insufficient \ion{Fe}{2}\,High opacity between 2650 and 2700\AA\/, while also
producing too much higher velocity \ion{Fe}{2}\,Low.  This might occur if the
distribution is something different than a power law.  

This model suggests that the outflow's most important inhomogeneity is a narrow component near $-3,200\rm \, km\, s^{-1}$ that has a low covering fraction and
very high gas column (equivalently, high optical depth), primarily evidenced by the $\sim 2960$\AA\/ lines that appeared between the SDSS observation in 2005 and this KPNO observation in 2011. If this component is part of the main outflow, it should have a similar distance from the central engine. So, perhaps the inhomogeneity consists of new higher density condensations that have developed inside the lower optical
depth, higher covering fraction outflow. A higher density in the condensations would yield a lower
ionization parameter, but \ion{Mg}{2} and \ion{Fe}{2} ionic column density
contours run parallel to the ionization parameter (e.g.,
Fig.~\ref{FIG7}), so the ions present would
not necessarily change much. 

Is this component really part of the main outflow? There is some, rather weak, evidence in favor of this hypothesis.  Certainly, the presence of high-excitation iron in
the condensations requires a high density, and therefore an absorption
radius, like that of the broad line outflow, rather close to the
nucleus (\S\ref{comparison}).
Furthermore, this component varies on relatively short time scales
(i.e., between the SDSS observation in 2005 and the KPNO observation
in 2011), and timescales can be shorter closer to
the nucleus due to higher densities / shorter recombination times, or shorter crossing times for clouds moving across the line of sight \citep[e.g.,][and references therein]{capellupo11}. Furthermore, as discussed in \S\ref{new_lines}, there are
some hints that the velocity of these new cores was predicted, to some
extent, by the structure of the \ion{He}{1}* BALs in older spectra,
when the cores were not observed.

This analysis introduces a note of caution into our
constraints; the presence of dense, low covering fraction
inhomogeneities throws some uncertainty into our assumption that the
spatial gas distribution can be approximated as a step function. In
particular, additional mass may be hidden in such cores, which we do
not consider in the following section because it cannot be constrained
with the available data; for example, the new lines appeared sometime between
the 2005 and 2011, so the upper limit on
density obtained from Balmer constraints in \citet{leighly11} (using the 2005 spectrum) may not apply to the cores.

\subsection{A Density Lower Limit\label{density}}

To obtain a lower limit on the characteristic number density in the BAL flow, we investigated the dependence
of Fe$^+$ level populations on density and $\log(N_H)-\log(U)$ for
fixed $\log U=-1.5$.  We first looked at the dependence for
\ion{Fe}{2}\,Low,
finding that $\log n > 5.1$ is required to produce the measured column
density.     We then looked at  the density dependence of levels between
0.98 and $1.72\rm \, eV$, i.e, levels that contribute to both
\ion{Fe}{2}\,High and to the \ion{Fe}{2} near 2960\AA\/.  For column
densities up to $\log(N_H)-log(U)=24$, we do not populate the levels
near 1 eV and 1.7 eV sufficiently unless the density is greater than
$\log n=5.5$ and $\log n=6.2$, respectively.  It was the lack of the
broad component absorption in the 1.7 eV levels that contributed to our column density
upper limit of $\log(N_H)-log(U)=23.4$ (\S\ref{upper_limit}), so we instead rely on the 1 eV level constraint, yielding
$\log n=5.5$ as our conservative density lower limit.

\subsection{Kinematic Constraints\label{outflow_param}}

In this section, we describe new constraints on the kinetic parameters
of the outflow.  We followed the standard methodology, which was
described and used on this object in \citet{leighly11}.
As detailed  in \citet{leighly11}, we used a characteristic
column-weighted velocity of $-5400\rm\, km\, s^{-1}$, an assumed global
covering fraction $\Omega = 0.2$, and a matter-to-radiation conversion
efficiency of 10\%.  We made one small change, however; due to a
more careful dereddening presented in this paper, the estimated
optical flux is larger by 20\%.  This in turn increases the estimated
black hole mass by 4\%, to $8.5\times 10^8\, M_\odot$, and the
bolometric luminosity became $6.3\times 10^{46}\rm \, erg\, s^{-1}$
(obtained by integrating over the {\it   Cloudy} model continuum
scaled to the dereddened optical flux).  

\begin{deluxetable}{lcc}
\tablecaption{Kinematic Limits for the Broad Outflow}
\tablewidth{0pt}
\tablehead{\colhead{Parameter} &
\colhead{Weakest Outflow } & 
\colhead{Strongest Outflow }}
\startdata
log Ionization Parameter & $-1.5$ & $-1.5$ \\
log Density [$\rm cm^{-3}$] & $8.0$  & $ 5.5$ \\
$\log N_H - \log U$ & 23.18 & $23.4$ \\
$\log N_H$\tablenotemark{a} & 21.68 & 21.9 \\
\tableline
Radius [$\rm pc$] & 7.2 & 127 \\
Mass Flux\tablenotemark{a} [$\rm M_\odot\,yr^{-1}$] & 10.7 & 315 \\
Kinetic Luminosity\tablenotemark{a} [$\rm 10^{44}\, erg\, s^{-1}$] &
0.99 & 28.9 \\
Kinetic Luminosity\tablenotemark{a} / Bolometric Luminosity & 0.16\% & 4.5\%\\
Outflow Rate\tablenotemark{a} / Accretion Rate\tablenotemark{b} & 0.96 & 28.3 \\
\enddata
\tablenotetext{a}{\footnotesize Assuming global covering fraction $\Omega = 0.2$.}
\tablenotetext{b}{\footnotesize Assuming an accretion efficiency of 10\%.}
\label{table5}
\end{deluxetable}

We investigated the kinematic properties of the BAL-producing outflow,
characterized by broad \ion{Fe}{2}\,Low, \ion{Fe}{2}\,High,
\ion{Mg}{2}, and \ion{He}{1}*, ignoring the kinematics of poorly
constrained dense-core inhomogeneities posited in the preceding section. In \citet{leighly11}, we had to assume acceleration by radiative line driving (which yielded a density lower limit) in order to constrain kinematic properties; the additional information provided here by \ion{Fe}{2} has allowed us to lift that assumption. \S\ref{fom}, \S\ref{upper_limit}, and
\S\ref{density}, along with a density maximum obtained from upper limits on Balmer
absorption in \citet{leighly11}, constrained the gas
parameters in the main outflow to the following ranges: $\log U
= -1.5$, $5.5 \leq \log n \leq 8$, $23.18 \leq \log N_H - \log U \leq
23.4$, and correspondingly $ 21.68 \leq \log N_H \leq 21.9$. Choosing
combinations of these limits that extremize the outflow's power and radius, we
constrained kinematic parameters to the following ranges: radius
(distance from the central engine) between 7.2 and 127 parsecs, mass flux
between 10.7 and 315 $\rm M_\odot\,yr^{-1}$, and kinetic luminosity
between $1.0\times 10^{44}$ and $29 \times 10^{44}$ $\rm erg\,
s^{-1}$. This range of kinetic luminosity is between 0.16\% and 4.5\%
of the bolometric luminosity, and the ratio of mass flux to accretion
rate is between 0.96 and 28.3. These constraints are summarized in
Table~\ref{table5}.

Our results depend somewhat on the  shape of the input                                                                                        
  SED, which we cannot estimate for FBQS~J1151$+$3822; the UV  continuum appears to be reddened, and there is no X-ray detection.  We used a generic AGN continuum from \citet{korista97}\footnote{\footnotesize The {\it Cloudy} command for this continuum
  is ``AGN Kirk.''}, which we have used with success on various objects, and on this object in \citet{leighly11}.  But to investigate the influence of SED choice,
  we have performed some simulations using the much softer SED that
  \citet{hamann13} use. This SED would yield an inferred distance from the continuum
smaller by $\sim 28$\%; this is negligible considering the range of uncertainties.
  The softer SED is, however, so weak in X-rays that it
  produces a cool partially ionized zone, increasing the upper limit
  on $\log(N_H)-\log(U)$ to greater than 24.  Such a large amount of
  material (i.e., $\sim 10$ times the Stromgren sphere thickness)
  might be difficult to accelerate.  At any rate, given that
  we have no direct information about the shape of the SED, and for
  continuity with our previous paper, we assumed the generic
  \citet{korista97} continuum.

\section{Discussion\label{discussion}}

\subsection{Inhomogeneous Absorption in FBQS~J1151$+$3822\label{inhomogeneous}}

The simplest interpretation of partial covering, that part of the
  continuum source is covered uniformly while the remainder is
  uncovered, may not be generally valid. Rather, the distribution of
  hydrogen column can be inhomogeneous across the line of
  sight. \citet{hamann01} propose an absorber of large/low-density
  blobs that allow for sufficient opacity from strong (but not weak)
  lines to be observed, along with small/higher-density cores that
  allow for sufficient opacity from weak lines. In this model, the
  observed step-function covering fractions are larger for stronger
  lines, matching their observations. Lines with small oscillator
  strengths and/or low species abundance only sample the smaller cores of
  dense gas, which have less overall coverage of the source. That is,
  weak lines will show smaller covering fractions when one applies the
  step-function covering fraction model to a very inhomogeneous gas. 

Aside from that, lines that arise in different physical
  conditions can also have different covering fractions. A related
  idea, similar to the scenario proposed by \citet{voit93}, might
  have radiation pressure ablating extended
  lower-density/higher-ionization gas from dense cores; this picture
  has the advantage that it could also explain the extension of
  higher-ionization lines to larger velocities observed in some objects.
  
  \citet{arav05, arav08} have explored some simple
inhomogeneous distributions, such as a power law distribution in one
spatial dimension.  In \citet{arav08}, homogeneous and inhomogeneous
fits yield ionic columns within 25\% of each other, and similar parity
between methods was seen in \citet{leighly11}. However, extreme power laws, not
to mention more complex distributions, could behave less like a
step function distribution, and there appears to be little observational evidence in the literature to suggest a universal distribution. 

As discussed in \S\ref{plausible}, inhomogeneous partial covering is required to
explain the spectrum of FBQS~J1151$+$3822.   Expressed in terms of a
geometrical covering fraction, we find that the spectrum can be fit
best if the line-of-sight covering fractions around $\sim 0.25$ for
\ion{Fe}{2}, increasing to $\sim 0.5$ for \ion{He}{1}* and $\sim 0.6$
for \ion{Mg}{2} (e.g., Fig.~\ref{FIG3}, Fig.~\ref{FIG9}).  In
addition, the lowest opacity lines, exemplified by the new, relatively
narrow lines observed near 2960\AA\/, may have an even lower covering
fraction in the vicinity of $\sim 0.1$ to $\sim 0.2$. 

We can use this information, combined with the information in
Fig.~\ref{FIG10}, to crudely map the column density to the covering
fraction. For example, we know that gas with $\log N_H-\log U\approx
23.2$, where \ion{Fe}{2} saturates,  must cover less than $\sim 50$\%
of the continuum emission source, otherwise the \ion{Fe}{2} lines
would be as deep as the \ion{He}{1}*$\lambda 10830$ and \ion{Mg}{2}
lines. Likewise, gas with $\log N_H - \log U \ge  23.5$ must cover
less than $\sim 30$\% of the continuum emission source, or the lines
near 2960\AA\/ would be deeper.  Information from higher-ionization
lines, such as \ion{Si}{4}, which saturates at  $\log N_H - \log U
\approx 22.8$, and  \ion{C}{4}, which saturates at $\log N_H - \log U
\approx 22$, would fill in information for smaller column densities.

With more line species, a distribution of
column density as a function of effective covering fraction could be
made, under the assumption of constant density, that could
potentially be used to build a model of inhomogeneity.  This also brings
out a under-appreciated difficulty in analyzing BALQSO spectra; that
is, if the covering fraction can be measured using two lines
\citep[e.g., as was done in ][]{leighly11}, that covering fraction
often cannot be assumed to apply to other lines less amenable to
covering fraction analysis (e.g., if the gas is non-uniform {\it and} lines are not equally strong, or do not form in equal amounts in identical conditions). 

Finally, it is interesting that \ion{Mg}{2}$\lambda2800$ has, in our best estimates, a covering fraction somewhat higher than that of \ion{He}{1}*$\lambda10830$ ($\sim 0.6$ to $\sim 0.5$).  As discussed
in \citet{leighly11} and elsewhere, the continuum emission region is
expected to be extended, with the longer wavelength continuum produced
over a more extended area.  For example, for $M_{BH}=8.5\times 10^8 \,
\rm M_\odot$ and $\dot{M} = 11.1 \, \rm M_\odot\, yr^{-1}$, the
2800\AA\/ and 10830\AA\/ regions are expected to have peak emission
at $0.008$ and $0.047\rm \, pc$, respectively, from the
center, based on a sum-of-blackbodies accretion disk model \citep[e.g.,][]{frank02}.  Thus, it is possible that the projected size of the absorber
is large compared with the 2800\AA\/ emitting region, and slightly
smaller compared with the 10830\AA\/ emitting region. 

\subsection{Comparison with Other FeLoBALs \label{comparison}}

The unambiguous detection of \ion{Fe}{2} absorption reclassifies
FBQS~J1151$+$3822 as an FeLoBAL.  With $z=0.3344$ and $M_V=-25.5$, it
is presently the second-nearest and second-brightest example of this class, after Mrk 231.   Detailed analyses
of \ion{Mg}{2}, \ion{Fe}{2} and \ion{He}{1}* in  FeLoBALs have
been presented by other authors.  However, because this type of
analysis is very difficult, as a consequence of severe blending and
the large number of lines that may be present, it has been applied to
only a handful of objects.  Here we discuss those results, and compare
with our results from FBQS~J1151$+$3822.   This comparison is summarized
in Table~\ref{table7}, and the spectra of several objects are
presented in Fig.~\ref{FIG12}.   

In many cases, the authors
used a template method to analyze the data, 
like we did; they generally developed the template from a
narrow, unblended line from e.g., \ion{Mg}{1}, \ion{Fe}{2}, or
\ion{Zn}{2} \citep{dekool01,dekool02a,dekool02b}.  In some objects,
the lines were sufficiently narrow and unblended that a direct
line-fitting approach could be used \citep{arav08, moe09, dunn10}.  

In some cases,  the presence  of lines determined to originate in
well-populated, highly-excited \ion{Fe}{2} led the authors to conclude
that the gas is dense, with densities greater than, e.g.,  $\log n_H =
10^{6}\rm \, cm^{-2}$, and the absorption region compact, lying
within  e.g., 1 pc from the central engine 
\citep{wampler95,dekool02a,dekool02b}.  In other cases, the
high-excitation \ion{Fe}{2} lines are absent, and the low-excitation
\ion{Fe}{2} displays ratios that suggested low densities, and therefore large
distances between the absorbing gas and the central engine, e.g.,
hundreds to thousands of parsecs \citep{dekool01, dekool02b, 
korista08, moe09, dunn10}.  In one case, the absorption was separated
into two kinematic components, with the high-velocity lines consistent
with high-density gas, and the low-velocity lines consistent with
low-density gas \citep{dekool02b}.   

Several authors concluded that \ion{Fe}{2}
is produced in a thin layer near the hydrogen ionization front
\citep{wampler95, dekool01, korista08}.  Some speculated that
the width of that region could be broadened by the presence of dust
\citep{dekool01}, while in other cases dust is ruled out by the
absence of strong neutral absorption \citep{dekool02a}.  Some have
commented on the fact that \ion{Mg}{2} is deeper than \ion{Fe}{2},
indicating a higher covering fraction \citep{wampler95}, while others
have noted, like we have, that the most prominent \ion{Fe}{2} lines are saturated
\citep{dekool01}.

\begin{figure}[p]
\epsscale{0.7}
\includegraphics[width=4.0in]{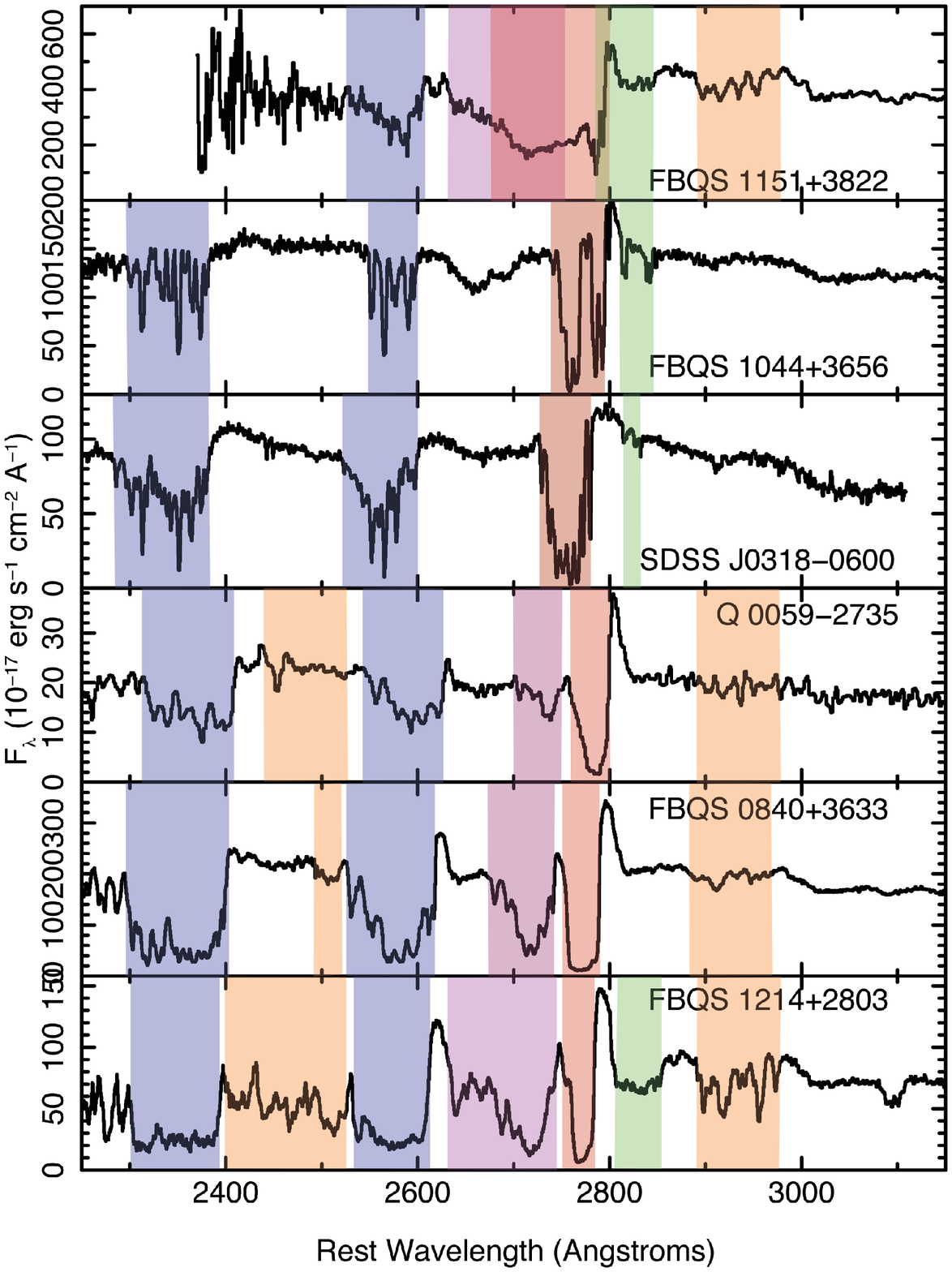}
\caption{\footnotesize FBQS~J1151$-$3822 in comparison with spectra of
  other bright, well-studied FeLoBALs, including FBQS~J1044$+$3656
  \citep{dekool01}, SDSS~J0318$-$0600 \citep{dunn10}, Q~J0059$-$2735
  \citep{wampler95}, FBQS~J0840$+$3633 \citep{dekool02b}, and
  FBQS~J1214$+$2803 \citep[e.g.,][]{dekool02a}. Note that the Q~J0059$-$2735 
spectrum is a lower-resolution, lower signal-to-noise ratio spectrum
published by \citet{weymann91}, and thus some of the
low-equivalent-width line identifications are uncertain.    Shaded
regions denote the  extent of various absorption lines, and
absorption-line classes, as follows: red: \ion{Mg}{2}$\lambda\lambda
2796,2803$,  observed near 2770\AA\/; green: \ion{Mg}{1}$\lambda
2852$, observed near 2830\AA\/; blue: low-excitation \ion{Fe}{2},
observed between $\sim 2300$ and $\sim 2400$\AA\/ and between $\sim
2530$ and $\sim 2600$\AA\/; purple: high opacity high-excitation
\ion{Fe}{2}, observed between $\sim 2650$ and $\sim 2750$\AA\/;  
orange: low-opacity high-excitation \ion{Fe}{2}, observed between
$\sim 2900$ and $\sim 2970$\AA\/, and between $\sim 2400$\AA\/ and
$\sim 2520$\AA\/ .  Objects with low-density absorbing gas,
including FBQS~J1044$+$3656 and SDSS~J0318$-$0600, can be easily
distinguished by eye (by their lack of high-excitation \ion{Fe}{2}
emission) from objects dominated by higher density gas,
including FBQS~J1151$+$3822, Q~J0059$-$2735, FBQS~J0840$+$3633 and
FBQS~J1214$+$2803.\label{FIG12}}     
\end{figure}

Fig.~\ref{FIG12} shows a comparison of our KPNO spectrum of
FBQS~J1151$+$3822 with the SDSS spectra of FBQS~J1044$+$3656,
SDSS~J0318$-$0600, FBQS~J1214$+$2803, and FBQS~J0840$+$3633, and a
spectrum of Q~J0059$+$2735 previously published by \citet{weymann91}.  
On this plot we mark the  extent of lines from various 
ions including \ion{Mg}{2}, \ion{Mg}{1}, and \ion{Fe}{2}, breaking the
\ion{Fe}{2} into low-excitation \ion{Fe}{2} (rest wavelengths 2586 -- 2631\AA\/ and 2327 -- 2411\AA\/, lower
level energies 0 -- $0.12\rm \, eV$), high-opacity
high-excitation \ion{Fe}{2} (2693 --
2773\AA\/; 0.98 -- $1.1\rm \, eV$),
and low-opacity high-excitation \ion{Fe}{2} (2372 -- 2641\AA\/, 2944 -- 3002\AA\/; 1 -- 3 eV, plus 1.17 eV \ion{Mn}{2}).  

A general sense
  of physical conditions in the  
absorbing gas can be gained by simply examining Fig.~\ref{FIG12} to see which types of
\ion{Fe}{2} lines are present and how deep they are.  For example,
FBQS~J1044$+$3822 has no \ion{Fe}{2} High, and therefore it can be
concluded on-sight that the
density is low and excited states are underpopulated.  
SDSS~J0318$-$0600 is similar.  FBQS~J1214$+$2803, on the other hand,
shows absorption lines from ions with very high excitation and high column density, indicating high density.
The same goes for FBQS~J1151$+$3822, which may also have the broadest absorption features.

\begin{deluxetable}{lccl}
\tablewidth{0pt}
\tablecaption{FeLoBAL Parameter Comparison}
\tablehead{
 \colhead{Target} &
 \colhead{Inferred Density} & 
 \colhead{Inferred Distance} &
 \colhead{References} \\
& \colhead{[$\rm cm^{-3}$]} & [pc]}
\startdata
Q0059$-$2735 & $10^6$--$10^{8.5}$ & 0.6--40 & \citet{wampler95} \\
FBQS J1044$+$3656 & $\sim 4\times 10^{3}$ & 700 & \citet{dekool01} \\
FBQS~J1214$+$2803 & $> 10^6$ & 1--30 & \citet{dekool02a} \\
FBQS J0840$+$3633\tablenotemark{a} & $10^9$ -- $3\times 10^{10}$ &
1--30 &  \citet{dekool02b} \\ 
& $< 500$ & $\sim 230$ \\
QSO J2359$-$1241 & $10^{4.4}$ & 3,000 & \citet{korista08} \\
SDSS J0838$+$2955 & $10^{3.75}$ & 3,300 & \citet{moe09} \\
SDSS J0318$-$0600 & $10^{3.3}$ & 6,000-17,000 & \citet{dunn10} \\
FBQS J1151$+$3822 & $10^{5.5}$--$10^{8}$ & 7.2--127 & this paper \\
\enddata
\tablenotetext{a}{\footnotesize \citet{dekool02b} report that there are two
kinematically-distinguishable components in this object that are
characterized by different densities and distances from the central engine.}
\label{table7}
\end{deluxetable}
Table~\ref{table7} 
gives estimated densities and distances of the
absorber from the central continuum source for various FeLoBALs, as given in the referenced
publications.  These trends substantiate our assertions from the
figure;  the two objects that lack \ion{Fe}{2} absorption from highly
excited iron ions have low estimated densities, and the distances from
the central engine are large, while the objects with strong
high-excitation \ion{Fe}{2} have high estimated densities, and the
distances from the central engine are small.  Note that it is somewhat
difficult to compare the distances directly, since these objects have
a range of intrinsic luminosities.  As will be discussed in \S\ref{den_sen},
it would be easier to compare distances if they were scaled by a
characteristic radius that accounts for the variations in intrinsic
luminosity.  Nevertheless, here we use absolute distance in parsecs
for comparison with previous work.

\citet{dekool02b} have already suggested that there are,
among \ion{Fe}{2} absorbers, perhaps two classes with different
characteristics: one with lines consisting of ``one or several narrow
components, sometimes blended together into wider structures that are
formed in a low-density gas at a large distance (0.1--$10\rm\, kpc$)
from the nucleus," and another with broader lines and broader substructure with densities ``higher by
several orders of magnitude [. . .] best explained if they are 1000 times closer
to the nucleus, 0.1--$10\rm \, pc$.''  The radius constraints for the FBQS~J1151+3822 outflow , at 7.2 to 127 pc, straddle the gap between de Kool's classes. This could in part be because our constraints are very conservative, because de Kool et al.'s analysis was based on only a few targets, or because they did not account for differing black hole masses.

Moreover, our {\it Cloudy} models suggest that the bimodality de Kool et al. see in FeLoBAL spectra is not, in fact, {\it necessarily} accompanied by a physical bimodality.  Fig.~\ref{FIG13} shows $I/I_0$ for \ion{Fe}{2}
Low and \ion{Fe}{2} High as a function of density and
$\log(N_H)-\log(U)$ for an example ionization parameter of $\log
U=-1.5$ and $\Delta V=5,000 \rm \, km\, s^{-1}$.  Examining this
figure, we see the expected decrease in $I/I_0$ at the hydrogen
ionization front near $\log(N_H)-\log(U)=23.2$.  But we also see that
while \ion{Fe}{2} Low 
absorption is strong enough to saturate the line at all densities,
\ion{Fe}{2} High saturates only for densities higher than $\log(n)
\sim 5.5$, and that \ion{Fe}{2} High goes from being negligible to
saturated over a relatively small range in density.  

This behavior,
plus the rapid transition to saturation present just past the hydrogen 
ionization front,  means that for a population of objects with a flat or unimodal distribution of characteristic densities, one is still likely
to see either objects with \ion{Fe}{2}\,Low alone
(low density objects with hundreds-pc to kpc scale absorbing regions) or objects with
both strong or saturated \ion{Fe}{2}\,Low and \ion{Fe}{2}\,High (high density objects with pc to tens-of-pc scale absorbing regions).  Thus,
we find the bimodality of \ion{Fe}{2} absorption in quasars that
\citet{dekool02b} observed can be reproduced from the astrophysics of \ion{Fe}{2} absorption, without recourse to a physically bimodal population of FeLoBAL quasars.

\begin{figure}[h]
\epsscale{1.0}
\includegraphics[width=5.0in]{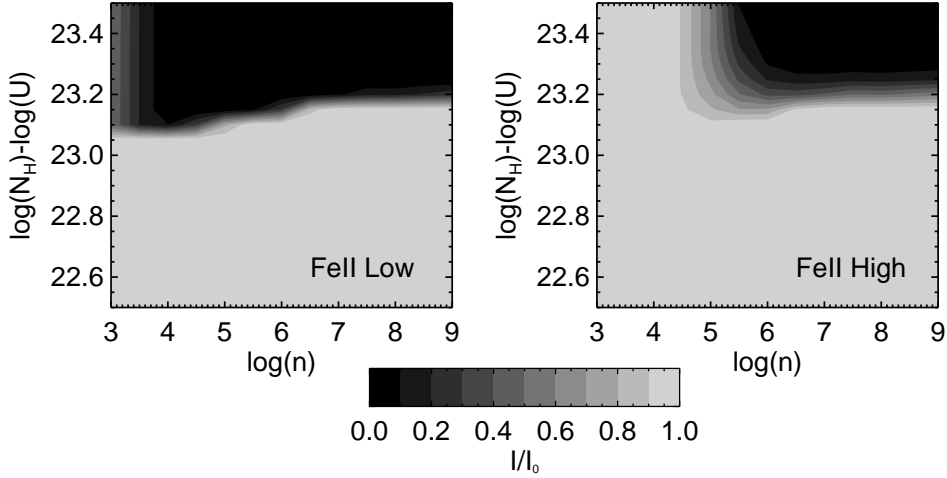}
\caption{\footnotesize Predicted $I/I_0$ for \ion{Fe}{2} Low (transitions with
  $E_{lower}$ near 0~eV, between 2500 and 2650\AA\/) and
  \ion{Fe}{2} High (transitions with $E_{lower}$ near 1~eV,
  between 2650 and 2800\AA\/)    for $\log U=-1.5$ as a function $\log
  n$   and  $\log N_H - \log U$.   The absorber is assumed to fully
  cover the source and the   absorption profile is rectangular with a
  width of $5,000\rm \,   km\, s^{-1}$.  These contours illustrate the
  rapid onset of   \ion{Fe}{2} absorption near the hydrogen ionization
  front, and the sharp density dependence of \ion{Fe}{2} High. \label{FIG13}} 
\end{figure}

Finally, we note that this behavior of \ion{Fe}{2} could bias target selection in
the near-UV.  Density can be estimated with precision if ground and
excited state transitions of iron are present.  Detailed analysis is
easier and less ambiguous when lines are rather narrow and unblended.
Gas with ions populated to higher levels will produce more lines,
causing more blending.  So objects with lower-density gas and
therefore fewer lines would present spectra more amenable to
analysis.  Thus, objects with low-density gas, and therefore kiloparsec-scale
outflows, might be  preferentially selected.

\subsection{Kinematic Comparison to Other
  BALQSOs\label{den_sen}}

Recently, some authors have made very speculative statements about locations of the BAL gas \citep[e.g.,][]{borguet13,arav13}. While perhaps implicitly allowing for a population of outflows $\sim 100$ pc away from the continuum source \citep[e.g.,][]{borguet13}, these authors strongly emphasize powerful outflows at kiloparsec scales, with $\sim 100$ pc seeming like a rather strong minimum. For example, \citet{arav13}, who use \ion{O}{4}/\ion{O}{4}* as a density diagnostic, say that their absorption radii are ``4+ orders
of magnitude farther away than the assumed acceleration region (0.03 -- 0.1 pc) of line-driven winds in quasars. This result is consistent with almost all the distances reported for AGN outflows in the literature."

  In contrast, we  find that the BAL outflow in
FBQS~J1151$+$3822 is located, conservatively, somewhere between 7.2 and 127 pc
from the continuum source; our upper limit overlaps only with the extreme end of the Arav group's range of objects, and most of our allowed radii range is on the tens-of-parsecs scale.  Does FBQS~J1151$+$3822 then have an unusually low-radius BAL outflow?

 We think not; rather, we hypothesize that the many AGN outflows in the literature could be subject to a selection bias.  These quasars were chosen
because their spectra contain lines from ground and excited state ions
with ratios that can be used to estimate the density.  As we will
show, this selection preferentially picks out objects with low
densities and therefore large absorption radii.

Consider, for example, \ion{S}{4}$\lambda 1063$ and
\ion{S}{4}*$\lambda 1073$ \citep{leighly09,borguet13}. These two lines represent transitions from the $^2P^0\,
J=1/2$ ground state and the $^2P^0\, J=3/2$ excited state.  The
critical density for the excited state line for a typical nebular
temperature of $10,000\rm \, K$ is $n_{crit}=4.2\times 10^{4}\rm \,
cm^{-3}$.  The density ratio of the first excited state to the ground
state will be equal to the column density ratio, assuming uniform gas
conditions (Fig.\ref{FIG14}). Similar plots are found in
\citet{borguet13} for \ion{S}{4}, for \ion{O}{4} in \citet{arav13},
for \ion{Fe}{2} in \citet{korista08}, and for \ion{Si}{2} in
\citet{dunn10}. 

To examine the useful range of density over which this line ratio can
be used, we consider some limiting cases.  We ignore partial
covering, which can add considerable uncertainty to
the line measurement.  As density increases, the excited state will
become more populated.  We assume that we can measure absorber populations when their observed lines are at least as deep as $I/I_0(v)=0.95$, and at least as shallow as $I/I_0(v)=0.05$.   Likewise, the highest density we can
measure with a reasonable degree of accuracy requires that neither
line be saturated, and therefore they should have different values of
$I/I_0$.  Since the excited state can have a higher population than
the ground state for this pair of transitions, we assume that it has
$I/I_0(v)=0.05$ when the ground state has $I/I_0(v)=0.1$.  Using these
two limits shows that the useful density range for this line is
between $4.5\times 10^2 \rm \, cm^{-3}$ and $1.7 \times 10^5\rm \,
cm^{-3}$, or between 0.01 and 3 times the critical density.
The precision with which these measurements can be made depends on the
quality of the data; nevertheless, it is clear that any particular
density diagnostic works best in the vicinity of the critical
density.

\begin{figure}[h]
\includegraphics[width=6in]{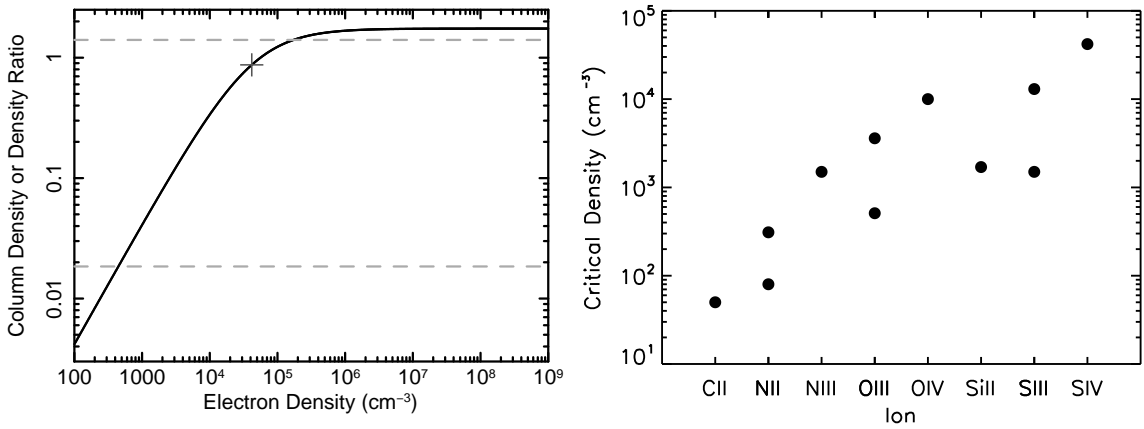}
\caption{\footnotesize {\it Left:} The predicted ratio of density of the excited
  state   \ion{S}{4}$\lambda 1073$ to the density of the ground state
  \ion{S}{4}$\lambda 1063$, calculated assuming a temperature of 
  $10,000\rm\, K$.  The dark gray plus sign denotes the critical
  density ($4.2\times 10^4\rm \, cm^{-3}$).  The dashed lines show the
  plausible upper and lower   limits for secure ratio 
  measurements (see text for details).  These imply that \ion{S}{4}
  can be used to measure   densities between $4.5\times 10^2 \rm \,
  cm^{-3}$ and $1.7 \times   10^5\rm \, cm^{-3}$, or between 0.01 and
  3 times the critical   density, suggesting a selection bias inherent in using density-sensitive lines. {\it Right:} Critical densities for
  common ions.  Ions with a single value of critical density have two
  levels of fine structure in their ground state.  Ions with two
  values have three levels, hence different values of critical density
  for each of the two excited states. \label{FIG14}}  
\end{figure}

Fig.~\ref{FIG14} shows the critical densities for many of the most
common ions with ground-state fine structure that
are seen in BAL outflows.  The first problem is that ions with
three-level ground states with $J=0,1,2$ have two critical densities,
from the $J=1$ and $J=2$ levels.  These critical densities can be
dramatically different, making these lines particularly difficult to
analyze since the separation of the lines is small.  This
issue was discussed for metastable \ion{C}{3}*$\lambda 1175$ observed
in NGC~3783 by \citet{gabel05}.  The second problem is that the
density-sensitive lines have a limited range of critical densities,
all less than that of \ion{S}{4}.  Since the \ion{S}{4} can be used to
securely measure densities less than $\sim 1.7\times 10^5\rm \,
cm^{-2}$, as we have shown above, selecting objects with lines
amenable to this kind of analysis pre-selects objects that have low
outflow densities. 

How does this selection impact the inferred radii?  The radius is
linked to the density through the ionization parameter and the
ionizing flux.  For a range of ionization parameters and densities, we
can compute the radius for a fixed number of ionizing photons.
It first makes sense to normalize the radius by dividing by a
characteristic radius, to account for the wide range of possible AGN luminosities.  We normalize by $R_{TK}$, the
reverberation mapping radius of the K-band photometry, which can be
taken to be the radius of the torus.  $R_{TK}$ is given by
\citet{kishimoto11} as $R_{TK}=0.47 (6\nu
L_{\nu}(5500$\AA\/$)/10^{46}\rm \, erg\, s^{-1})^{1/2} \,pc$, where
$6\nu L{\nu}(5500$\AA\/) is a surrogate for the UV luminosity.  For 
FBQS~J1151$+$3822, $R_{TK}=0.82\rm \, pc$, for the low-luminosity
BALQSO WPVS~007 \citep{leighly09}, $R_{TK}\approx 0.023\rm \, pc$, and
for HE0238$-$1904 \citep{arav13}, $R_{TK}\approx 1.3\rm \, pc$.  Thus,
for typical BALQSOs, $R_{TK}$ is on the order of a parsec.

We plot the inferred outflow radii normalized to $R_{TK}$ in
Fig.~\ref{FIG15} as a function of a typical range of ionization
parameters and density.  We also mark $1.7 \times 10^{5}\rm \,
cm^{-3}$, the plausible upper limit of density obtainable using the
density-sensitive lines.  This shows that, for a typical ionization
parameter of $\log U=-1.5$, no distances smaller than 
$R_{out}/R_{TK}=210$ would be measurable.  That value decreases to 67
for a higher ionization parameter of $\log U = -0.5$ and increases to
667 for a lower ionization parameter of $\log U = -2.5$.  More typical
densities reported in the literature, e.g., $\log n \approx 4$, yield
a lower limit on $R_{out}/R_{TK}$ between 275 and $2750 \rm \, pc$,
i.e., very far from the central engine.  Since radius enters linearly
in the equation for kinetic luminosity \citep[e.g., Eq\ 1 in
][]{arav13}, a large radius inevitably produces a large kinetic
luminosity. 

\begin{figure}[h]
\includegraphics[width=6in]{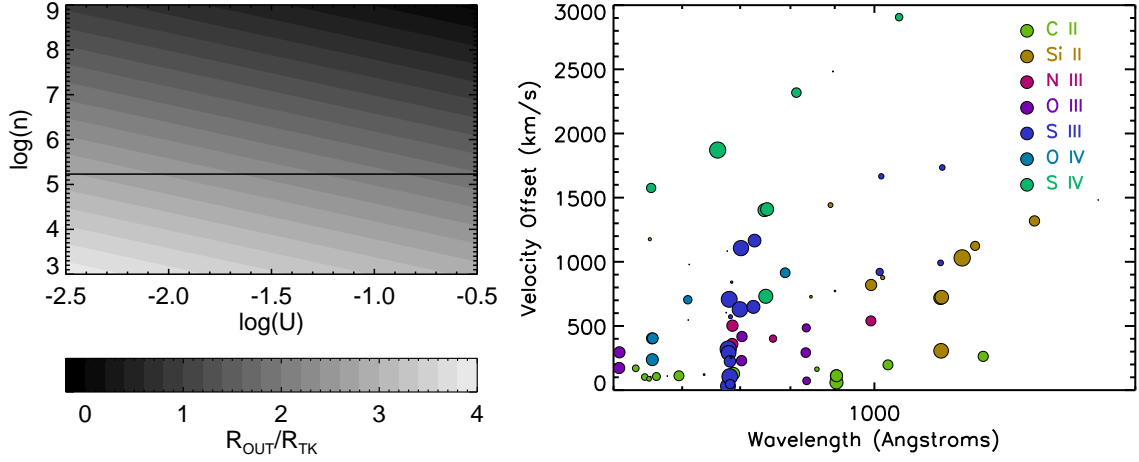}
\caption{\footnotesize {\it Left:} The ratio of the outflow radius $R_{out}$ to the
  K-band reverberation radius $R_{TK}$ as a function of ionization
  parameter and hydrogen density.  The solid line denotes the
  electron density $n_e=1.7\times 10^5\rm \, cm^{-3}$, which, as
  discussed in the text, is plausibly the largest density that can be
  securely measured using density-sensitive lines.  Thus, only
smaller densities and correspondingly larger radii can be explored
using density-sensitive lines.  {\it Right:}  The velocity separation of
adjacent lines with differing lower levels from a single ion as a
function of wavelength of the shorter-wavelength line of each line
pair.  The symbol size is proportional to the log of 
the oscillator strength for the shorter-wavelength line of any line
pair.  This plot shows that velocity separations can be very small,
due to the close spacing of the energy levels, and that the short
wavelength region of the spectrum can be very crowded, and therefore,
blending can be significant.  \label{FIG15}}  
\end{figure}

Blending can also significantly limit the use of density-sensitive
lines. The broad absorption lines in FBQS~J1151$+$3822 are
significantly blended, and analysis of those lines requires
assumptions that inevitably degrade the precision of the results.  Since narrower lines will be less blended, 
objects studied using density-sensitive lines are often chosen on the
basis of line width.   
Density-sensitive line pairs are often closely spaced, since the
differences in the energies of the ground and excited state are
small. Fig.~\ref{FIG15} shows the velocity separation of
density-sensitive lines between 500 and 2000\AA\/ as a function of
wavelength. Some lines, such as \ion{S}{3}, have many transitions, and
so the lines are closely packed, making measurements of individual
lines difficult. Lines from different ions can be blended as
well. Note that we did not include resonance lines with
ground states originating in S terms, including, e.g.,
\ion{C}{3}$\lambda 977$, \ion{N}{4}$\lambda 765$, the
\ion{O}{5}$\lambda 630$, \ion{Si}{3}$\lambda 567, 1205$, and
\ion{S}{6}$\lambda 933, 945$; these lines would increase the degree of blending further.

Perhaps there are no objects with narrow lines showing large ratios of
excited state to ground state column densities (e.g., objects on the
high-density asymptote in Fig.\,\ref{FIG15}). But it may be that outflows
characterized by broader lines, which would not be attractive
candidates for density analysis, do have higher densities and originate
closer to the central engine.  This may make sense dynamically; if a BAL
outflow is a transient event \citep{leighly09, filizak12},
could it start near the nucleus with a high density, fragmenting,
dispersing, and evaporating as the distance increases? It may be
possible to test this idea by examining the \ion{S}{4} properties in,
for example, BOSS quasars. The \ion{S}{4} pair of lines, at 1062 and
1073\AA\/, has one of the widest separations available ($\sim 2900\rm
\, km\, s^{-1}$). There are few lines close to and longward of this
line pair, so the presence of the excited state can be clearly 
identified. \citet{leighly09} show that in WPVS~007, this line pair
could be modeled  using a template developed from the column-density
sensitive lines \ion{P}{5}$\lambda\lambda 1118, 1128$.  This could make sense because, although \ion{S}{4} is a slightly lower ionization
line compared with \ion{P}{5} and sulfur is about 100 times more
abundant than phosphorus, the oscillator strengths of the \ion{S}{4}
lines are about an order of magnitude lower than those of the
\ion{P}{5} lines.  So these two
ions might more likely be formed in a similar region of the outflow, and therefore
might be expected to share similar kinematics and covering fraction.
Finally, examining Fig.\ 6 in \citet{leighly09} shows that \ion{S}{4}*
is strong in both WPVS~007 and comparison object LBQS~1212$+$1445,
indicating high-density gas in both.   

Thus, we conclude that FBQS~J1151$+$3822, with its strong, blended
excited-state absorption and correspondingly high density and small
absorption radius, may not be genuinely unusual. Rather, the
application of a density analysis to this object may be what is
unusual.  We chose to analyze FBQS~J1151$+$3822 because we were
interested in finding out what the newly discovered \ion{Fe}{2} and
\ion{Mg}{2} absorption lines could tell us about the \ion{He}{1}*
absorption that we had previously reported, so it was not subject to this selection bias.

\section{Summary \label{summary}}

We obtained a KPNO spectrum of the \ion{He}{1}* BALQSO FBQS~J1151$+$3822
sampling short wavelengths approaching the atmospheric cutoff.  We
summarize our results below. 

\begin{itemize}
\item The KPNO spectrum revealed broad, deep absorption
centered near 2580 and 2750 \AA\/.   The
  shorter wavelength feature is mostly ground term \ion{Fe}{2}. The longer wavelength feature contains \ion{Mg}{2} and excited \ion{Fe}{2}. 
These results demand reclassification of FBQS~J1151$+$3822 as an
FeLoBAL; it is, to our knowledge, the second-brightest ($m_V = 15.60$), second-closest ($z = 0.334$) known FeLoBAL after Mrk~231. 

\item We measured apparent ionic columns (Table~\ref{table_ioncolumns}, \S\ref{analysis}) of \ion{Mg}{2}, \ion{Mg}{1}, and two groups of \ion{Fe}{2} excited levels (0--$0.12\rm \, eV$ and 0.98--$1.1\rm \, eV$, which occur in largely disjoint parts of the spectrum), using \ion{He}{1}* lines as templates for de-blending. The
absorption  model fit the spectrum well overall, although not in
detail. 

\item We used the measured ionic columns and the photoionization code
{\it Cloudy} to determine the physical properties of the absorbing gas.
The extremely useful combination of \ion{He}{1}*, produced in the \ion{H}{2}
region, and \ion{Fe}{2}, produced beyond the hydrogen ionization front,
pinpointed the ionization parameter  at $\log U=-1.5$ (\S\ref{prelim_cloudy}). The {\it Cloudy} parameter space demonstrated that \ion{Mg}{2} is saturated. We found that the \ion{He}{1}*, \ion{Fe}{2},
\ion{Fe}{2}, and \ion{Mg}{1} apparent columns were best matched at $\log   U=-1.5$, $\log n=7.2$, and $\log N_H-\log U =
23.18$, corresponding   to $\log N_H = 21.68$ (\S\ref{fom}).   However, the consequently predicted ion columns did not yield an acceptable fit to the \ion{Fe}{2} absorption spectrum, and we determined that  \ion{Fe}{2} is also saturated (\S\ref{saturation}), so that the best {\it Cloudy} fit provides only a lower limit on the hydrogen column density.

\item To determine an upper limit on the hydrogen column density, we found an
  upper limit on absorption by low-opacity excited-state (0.98--$1.72\rm \, eV$) 
   \ion{Fe}{2} lines and excited-state ($1.17\rm \, eV$) \ion{Mn}{2} near 2960\AA\/ (\S\ref{upper_limit}).  The resultant upper limit on
  $\log(N_H) - \log(U)$ is 23.4, corresponding to an upper limit on $\log(N_H)$ of 21.9. The presence of broad, strong excited state
  \ion{Fe}{2}  implies a conservative lower limit on the density of
  $\log n \sim 5.5$ (\S\ref{density}).   
  
\item Using these results, we obtained lower and upper limits on the
  kinematic parameters of the BAL-producing outflow (\S\ref{outflow_param}).  In contrast  to \citet{leighly11}, we were able to obtain constraints without making the unjustified assumption that the outflow is accelerated by radiative line driving.   We found that
  the outflow radius is 7.2 to 127 parsecs, the mass flux 10.7 to $315\rm \, M_\odot\, yr^{-1}$, and the kinetic
  luminosity is $1\times 10^{44}$ to $28.9 \times 10^{44}\,
  \rm erg\, s^{-1}$ (0.16 to
  4.5\% of the bolometric luminosity).

  \item We estimated covering fractions to be around 0.6 for \ion{Mg}{2}, 0.5 for \ion{He}{1}*, and 0.25 for high-opacity \ion{Fe}{2} (\S\ref{upper_limit}) with the differences yielding some preliminary insight into the geometric distribution of hydrogen column (\S\ref{inhomogeneous}). With more ions, similar future analyses might yield a constraint on the inhomogeneity of the gas.

\item  Narrow, rather shallow features appeared between the 2005 SDSS spectrum and the 2011 KPNO spectrum of this object in the $\sim 2900$
  to $\sim 2960$\AA\/ range, which we identified as transitions of
  \ion{Mn}{2} and high-excitation \ion{Fe}{2} with velocities of $\sim
  3,300\rm \, km\, s^{-1}$ (\S\ref{new_lines}).  These may signify
  condensation of higher-density clumps in the outflow (\S\ref{plausible}). Additional mass, and therefore energy, may be hidden in these cores.
  
  \item Our analysis clarified the physical differences between HiBALs, LoBALs, and FeLoBALs (\S\ref{saturation}). For a given ionization parameter, HiBAL outflows have the smallest column, LoBAL outflows have a larger column (but \ion{Mg}{2} LoBALs can still be truncated before the hydrogen ionization front is breached), and FeLoBAL outflows have the largest column (thicker, at least marginally, than the \ion{H}{2} zone).
    
  \item In comparing our object to other well-studied FeLoBALs (\S\ref{comparison}), we found that the presence of \ion{Fe}{2} 0.98--$1.1\rm \, eV$ absorption indicates absorbing gas with density higher than critical and a distance from the nucleus on the order of tens-of-parsecs, while the absence of said absorption indicates absorbing gas with density lower than critical and a distance from the nucleus on the order of kiloparsecs. This absorption complex saturates over a small differential of physical parameter space, so a bimodality in this regard of FeLoBAL spectra does not necessarily indicate the bimodality of FeLoBAL physical properties posited in the literature. Thus, our object might not belong to a distinct low-radius subclass.
  
\item The FBQS~J1151$+$3822 outflow has higher density, and consequently
  smaller distance from the central engine, than most other intensively studied
  quasars---but this may be due to a selection bias in the literature.  We investigated the use of density-sensitive lines, which have been used to conclude most outflows are located on kiloparsec scales.  These lines have relatively low critical densities, and provide secure density estimates no higher than several times the critical density. Selecting for these lines, especially when unblended lines are preferred, may lead to a preference for large radii (\S\ref{den_sen}).

\end{itemize}



\acknowledgments

We thank Pat Hall, Dick Henry, Eddie Baron, and Martin Elvis for useful discussions, and acknowledge Sara Barber's participation in the KPNO observing run. A.B.L. acknowledges funding by the Barry Goldwater Scholarship Foundation, and with K.M.L. acknowledges funding by the National Science Foundation. We thank the referee for helping us improve the clarity of this paper. This 
research has made use of the NASA/IPAC Infrared Science Archive, which is 
operated by the JPL, Caltech, 
under contract with NASA.  
This publication makes use of data products from the Wide-field Infrared 
Survey Explorer, which is a joint project of UCLA and JPL / Caltech, funded by NASA.



{\it Facility:}  \facility{Mayall (R-C CCD Spectrograph)}



\appendix

\section{The Host Galaxy Contribution to the Near IR\label{host_gal}}

The one-micron region of an AGN or quasar spectrum is complicated.
The accretion disk (assumed to emit a power law spectrum), the host galaxy, and
the molecular torus can all contribute to varying degrees in this
region \citep{landt11}.  The BAL outflow is located at a smaller radius than the host galaxy, and 
therefore host galaxy light is unabsorbed by the BAL wind.  If host
galaxy emission is present in the spectrum coincident with the
\ion{He}{1}*$\lambda 10830$ absorption line, the covering fraction will be underestimated.   

The emission of the inner edge of the molecular torus, marked by an 
increase in flux toward longer wavelengths, is thought to be due to
hot dust at the sublimation radius.  The K-band reverberation radius, given by
$R_{TK}=0.47 (6\nu L_{\nu}(5500$\AA\/$)/10^{46}\rm \, erg\,
s^{-1})^{1/2} \,pc$ \citep{kishimoto11}, can be substituted for the
sublimation radius, and it 
is comparable with BAL outflow radii, generally speaking, so that
the torus emission may or may not be absorbed.  If it is not absorbed,
but remains in the spectrum, the covering fraction will again be
underestimated. 

We did not take the host galaxy emission into account when estimating
the \ion{He}{1}* column in \citet{leighly11}.  Here, we rectify that
oversight.  We estimate the host galaxy and torus contributions, and
then work through the remainder of the analysis presented in
\citet{leighly11}.  We show that, while the average opacity and
average covering fraction change when these components are accounted
for, the average column density remains the same.  

We estimated the contribution of the accretion disk and the torus to
the integrated light of FBQS J1151$+$3822 by modeling available
broadband photometry. This consisted of aperture magnitudes from the
Sloan Digital Sky Survey, 2MASS, and WISE.  FBQS J1151$+$3822 does not appear in
the Spitzer and UKIDSS catalogs.

We first correct the SDSS and 2MASS photometry for
emission and absorption lines.  We proceed by first merging the SDSS,
KPNO and IRTF spectra after having corrected for Galactic absorption.
The merged spectrum was analyzed in segments corresponding to each
filter.  A minimum and maximum plausible continua was identified in
each segment.   The estimated continuum and the spectrum are then folded with
the filter  transmission curve, and the ratio of these gives a
correction factor. The photometry was corrected
using the mean of the minimum and maximum correction factors.  One
half the distance between the minimum and maximum correction factors was
assigned as the uncertainty on the process of correcting for emission
and absorption lines, and this was combined in quadrature with the
photometry errors. We did not
apply any correction to the WISE photometry, and note that those long
wavelengths are less critical for determining the host galaxy
contribution.  

The individual magnitudes were converted to $F_\lambda$ units using
the calibrations in \citet{fukugita96} [SDSS], \citet{cohen03}
[2MASS], and \citet{jarrett11} [WISE], then corrected for Galactic
extinction using $E(B - V) = 0.024 \pm 0.004$ from \citet{sfd98}
and the optical/infrared reddening curve returned by IRAC, which is
based on $A_\lambda/A_V$ information in \citet{ccm98},
\citet{odonnell94}, and \citet{lutz96}. 

The line-corrected photometry was next corrected for Galactic reddening 
and then shifted to the rest frame.  The resulting photometry of the
continuum is shown in Fig.~\ref{FIG16}, along with the original
values.  The 
results were fit using Sherpa   \citep{freeman01} version 4.4 with a
continuum model consisting of a power law for the accretion disk with
$F_\lambda \propto \lambda^n$, a galaxy component
consisting of a 13 Gyr single-burst population from PEGASE models
\citep{fioc97}, a blackbody torus with temperature $T_t$, and a
blackbody cool dust component with temperature $T_d$; the latter only
contributes to the reddest two wavelengths from the WISE photometry.  We ran
a series of models in which the internal reddening for the AGN was
fixed at various values $E(B-V)$, as follows.  Quasar reddening
curves are often inferred to be steep, more like the SMC than like
Milky Way dust  \citep[e.g.,][]{crenshaw01}.  We implemented an SMC-like reddening function from
\citet{prevot84}.   The spectral fitting model was degenerate in the
power law slope, reddening, and galaxy template normalization.
Therefore, we fixed the reddening to two values: $E(B-V)=0.14$,
derived based on the shape of the blue part of the optical spectrum,
as derived above, and $E(B-V)=0.26$, based on the ratio of the Balmer
lines (observed to be 4.00) assuming an intrinsic ratio of 3.06
and an SMC reddening curve.  
While the Balmer decrement may seem to better estimate the
intrinsic reddening, note that it is not clear that the broad line
region is subject to the same reddening as the continuum, or that
the intrinsic ratio of the Balmer lines is 3.06
\citep[e.g.,][]{popovic03}.  Nevertheless, these two values provide
useful bounds.     

\begin{figure}[h]
\includegraphics[width=4in]{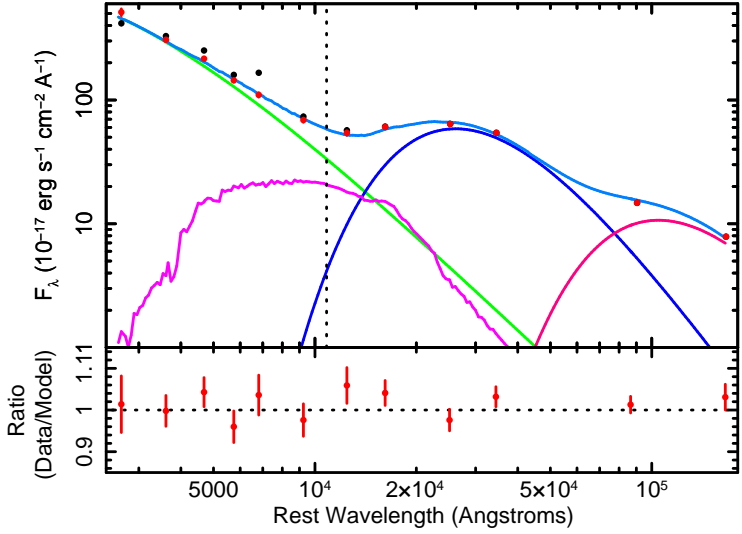}
\caption{\footnotesize The SDSS, 2MASS and WISE photometry fit with a model
  consisting of a power law (green line), two blackbodies (blue and
  red lines), and an elliptical
  template (fuchsia), all reddened by SMC-type extinction with $E(B-V)=0.2$
  (see \S\ref{host_gal}).  The black points give the raw
  photometry (corrected for Milky Way extinction and redshift), and
  the red points give the photometry corrected for emission or
  absorption lines in the bandpass.  At rest frame 10830\AA\/ (marked
  by the vertical line) the torus hot dust (i.e., warmer black body)
  and the host galaxy each contribute 7.1\% and 35.6\% of the
  continuum, respectively. This result is largely insensitive of the value of E(B-V) (see Table~\ref{table3}). \label{FIG16}}
\end{figure}

Table \ref{table3} displays the results of the fits to the
SED using these two values of $E(B-V)$, and for an intermediate value
of  $E(B-V)=0.20$, and the derived fraction of the contribution of the 
dusty torus and accretion disk to the continuum flux at $\lambda_0 =
10830$\AA\/. The derived slope of the accretion disk SED and the
starlight contribution of the disk to the continuum at the infrared
helium line are, as expected, dependent on the assumed internal
reddening; higher values of $E_0(B-V)$ require steeper power laws for
the accretion disk and result in a larger fraction of host galaxy.
The derived temperature of the AGN torus and the fraction of light
contributed by the torus are, however, largely insensitive to the
assumed extinction values.  If the reddening is within the assumed
range, then the contribution of the accretion disk and torus to the
continuum at $\lambda_{rest} = 10830$\AA\/ is $0.42 \pm 0.04$. Consideration of alternative extinction
SMC extinction laws such as \citet{pei92} and \citet{gordon03}
produced somewhat different values of the starlight/torus
contribution, which we characterize as an additional systematic error,
resulting in a final estimated contribution of $0.42 \pm 0.04$
(fitting) $\pm 0.06$ (systematic).   A plot of the fit for the
intermediate value of the extinction is shown in Fig.~\ref{FIG16}.

\begin{deluxetable}{lccc}
\tablewidth{0pt}
\tablecaption{SED fitting results}
\tablehead{
  \colhead{} &
  \multicolumn{3}{c}{$E_0(B - V)$} \\ \cline{2-4}
  \colhead{Quantity} &
  \colhead{0.14} &
  \colhead{0.20} &
  \colhead{0.26}\label{sed-results-tbl}
}
\startdata
Accretion disk slope    & $-2.33 \pm 0.015$ & $-2.59 \pm 0.013$ &
$-2.84 \pm 0.012$ \\
Torus temperature (K) & $1127 \pm 29$ & $1129 \pm 29$ & $1132 \pm 30$
\\
Torus fraction              & $0.073 \pm 0.003$ & $0.072 \pm 0.003$ &
$0.070 \pm 0.002$ \\
\
Starlight fraction          & $0.31 \pm 0.08$ & $0.35 \pm 0.08$ &
$0.39 \pm 0.09$ \\
$\chi^2_n$                   & 2.03 & 1.88 & 1.75 \\
\enddata
\label{table3}
\end{deluxetable}

We consider two cases: the
torus is absorbed by the outflow, and the galaxy is not, and both the
torus and galaxy are not absorbed.  We first create the ratio of the
galaxy (with and without the torus) to the total model, and transform
that to velocity space for a rest wavelength of 10830.171\AA\/.  We
resample that on the velocities derived from the original IRTF
spectrum \citep[shown in Fig.\ 3 in ][]{leighly11}.   We then subtract
this ratio from the profile and renormalize by dividing by one minus
the ratio.  The uncertainties assigned to the new $I/I_0$ ratios were
taken to be half the difference in the  values obtained by propagating
the original $I/I_0$ values plus and minus the original uncertainties.
As in \citet{leighly11}, we considered four profiles for the 3889\AA\/
line (obtained from two \ion{Fe}{2} models applied to the SDSS spectrum
and a later MDM spectrum); see \citet{leighly11} for details. 

The obtained average covering fraction ranged from 0.26 to 0.28 without continuum subtractions \citep{leighly11}, from 0.43 to 0.50 with the galaxy subtracted from the continuum, and from 0.48 to 0.51 with the galaxy and torus subtracted. The {\it average} column
density, which for the step-function partial covering model is equal to
the covering fraction times the column density of the covered
fraction, is much the same no matter whether we subtract the
galaxy, the torus plus the galaxy, or nothing; the
fit responds to the subtraction of continuum components by producing a lower average column density and higher
average covering fraction.  Therefore, we continue to use 14.9 as
the value the log of the \ion{He}{1}* column density, as we did in
\citet{leighly11}.




\small
\bibliographystyle{apj}
\bibliography{lucy}

\end{document}